\documentclass[a4paper,fleqn,usenatbib]{mnras}

\usepackage{newtxtext,newtxmath}

\usepackage[T1]{fontenc}
\usepackage{ae,aecompl}

\usepackage{graphicx}	
\usepackage{amsmath}	
\usepackage{amssymb}	
\usepackage{makecell}
\newcolumntype{C}[1]{>{\centering\arraybackslash}p{#1}}

\title[Radio-intermediate HERGs - Radio properties]{The radio properties of HERGs with intermediate radio powers}

\author[J. C. S. Pierce et al.]{
J. C. S. Pierce,$^{1}$\thanks{E-mail: jcspierce1@sheffield.ac.uk}
C. N. Tadhunter$^{1}$
and R. Morganti$^{2,3}$
\\
\\
$^{1}$Department of Physics and Astronomy, University of Sheffield, Sheffield S3 7RH, UK\\
$^{2}$ASTRON, The Netherlands Institute for Radio Astronomy, Postbus 2, 7990 AA, Dwingeloo, The Netherlands\\
$^{3}$Kapteyn Astronomical Institute, University of Groningen, PO Box 800, 9700 AV, Groningen, The Netherlands
}

\date{Accepted XXX. Received YYY; in original form ZZZ}

\pubyear{2019}

\begin{document}
\label{firstpage}
\pagerange{\pageref{firstpage}--\pageref{lastpage}}
\maketitle

\begin{abstract}
In the past decade, high sensitivity radio surveys have revealed that the local radio AGN population is dominated by moderate-to-low power sources with emission that is compact on galaxy scales. High-excitation radio galaxies (HERGs) with intermediate radio powers (22.5 $<$ log(L$_{\rm 1.4GHz}$) $<$ 25.0 W\,Hz$^{-1}$) form an important sub-group of this population, since there is strong evidence that they also drive multi-phase outflows on the scales of galaxy bulges. Here, we present high-resolution VLA observations at 1.5, 4.5 and 7.5 GHz of a sample of 16 such HERGs in the local universe ($z<0.1$), conducted in order to investigate the morphology, extent and spectra of their radio emission in detail, down to sub-kpc scales. We find that the majority (56 per cent) have unresolved structures at the limiting angular resolution of the observations ($\sim$0.3"). Although similar in the compactness of their radio structures, these sources have steep radio spectra and host galaxy properties that distinguish them from local low-excitation radio galaxies (LERGs) that are unresolved on similar scales. The remaining sources exhibit extended radio structures with projected diameters $\sim$1.4--19.0 kpc and a variety of morphologies: three double-lobed; two large-scale diffuse; one jetted and `S-shaped'; one undetermined. Only 19 per cent of the sample therefore exhibit the double-lobed/edge-brightened structures often associated with their counterparts at high and low radio powers: radio-powerful HERGs and Seyfert galaxies, respectively. Additional high-resolution observations are required to investigate this further, and to probe the $\lesssim$300 pc scales on which some Seyfert galaxies show extended structures.

\end{abstract}

\begin{keywords}
galaxies: active -- galaxies: jets -- galaxies: nuclei -- radio continuum: galaxies
\end{keywords}



\section{Introduction}
\label{sec:int}

The jets of radio AGNs are often deemed to affect the evolution of their host galaxies through two major feedback processes: i) driving multiphase outflows on the scales of galaxy bulges \citep[e.g.][]{mor05a,mor05b,holt08,mol19}; ii) preventing hot gas from cooling to form stars in galaxy haloes, or even the surrounding environment on larger scales \citep[e.g. see][]{mn07}. There is strong evidence to suggest that AGNs with intermediate radio powers (22.5 $<$ log(L$_{\rm 1.4GHz}$) $<$ 25.0 W\,Hz$^{-1}$) and high-excitation optical emission spectra (high-excitation radio galaxies, or HERGs) are particularly important for the former case, since they are seen to be associated with significant kinematic disturbances in both warm-ionised and molecular gas \citep{mul13,tad14a,har15,ram17,vm17}, and are considerably more common than their high-radio-power counterparts \citep{bh12}. Detailed characterisation of this population could therefore be crucial for providing a better understanding of the role of radio AGN feedback in galaxy evolution.

As a general population, the nuclear activity in HERGs has frequently been linked with the radiatively efficient accretion of cold gas at relatively high Eddington ratios \citep[e.g.][]{bh12,hb14,min14}. At high radio powers, HERGs are also predominantly associated with FRII-like (edge-brightened) radio morphologies, leading to suggestions that such structures could be inherently connected to this mode of accretion \citep[e.g.][]{tad16}. This is supported by the fact that some Seyfert galaxies, arguably the optically-selected equivalents of HERGs at low radio powers, are also seen to exhibit double-lobed or edge-brightened radio morphologies \citep[][]{uw84,mor99,baldi18b}.

However, the strength and nature of this connection between double-lobed radio structures and radiatively-efficient accretion across the range of radio powers covered by HERGs remains uncertain. For instance, \citet[][]{baldi10a} find that $\sim$80 per cent of a sample of local HERGs ($0.03<z<0.3$) selected from \citet[][]{best05a} have unresolved emission structures at the $\sim$5" limiting angular resolution of the FIRST survey \citep[][]{bec95}. This is also found to be true for 97 per cent of the 30 local HERGs with intermediate radio powers ($z < 0.1$; 22.5 $<$ log(L$_{\rm 1.4GHz}$) $< 24.0$ W\,Hz$^{-1}$) studied by \citet{pierce19}.

Moreover, several studies in the last decade have shown that the local radio AGN population as a whole is dominated by sources with radio emission that is compact on galaxy scales, given the resolution of typical radio surveys \citep[e.g.][]{baldi10a,sad14,whit16,baldi18a}. These sources are often collectively termed `FR0s', in order to separate them from the two traditional classes of extended radio structures: Fanaroff-Riley Type Is (FRIs) and Type IIs \citep[FRIIs, as above;][]{fr74}.

The majority of FR0s have massive, early-type host galaxies with large central black hole masses \citep{baldi15,baldi19a}, and typically exhibit low to intermediate radio powers \citep[10$^{22}$ $\lesssim$ L$_{\rm 1.4GHz}$ $\lesssim$ 10$^{24}$ W\,Hz$^{-1}$;][]{baldi18a}. They are in general identified with low-excitation radio galaxies (LERGs), in which the nuclear activity is thought to be fuelled by a radiatively-inefficient accretion flow at low Eddington ratios \citep[e.g.][]{bh12}. LERGs at high radio powers are commonly associated with FRI-like (edge-darkened) radio morphologies \citep{tad16}, and have properties that are consistent with the typical host galaxies of FR0 sources: they are also massive; of early-type; and have large black hole masses \citep[][]{baldi15,baldi19a,cap17}. In addition, FR0 sources with LERG spectra are found to extend the correlation between core radio luminosity and emission-line luminosity defined by FRIs down to lower radio powers \citep[][]{baldi19a}. 

This has led to suggestions that FRI and FR0 sources are members of a homogeneous LERG population, with the compact emission in FR0s being caused by either short cycles of radio activity or lower jet bulk speeds \citep[e.g.][]{baldi18a,baldi19a}. It is important to mention, however, that this picture is likely too simplistic, since these latter studies of FR0s are restricted to LERG sources only, and thus exclude known HERG sources with compact radio morphologies. For example, 2 out of the 11 (18 per cent) compact sources studied by \citet[][]{baldi15} have HERG spectra and radio structures that are compact on similar scales to their LERG/FR0s. In addition, a significant minority of high-radio-power LERGs have FRII-like, rather than FRI-like, radio structures \citep[e.g.][]{baldi10b,tad16}, which provide a further challenge to the idea of a homogeneous LERG population in terms of radio properties.

Although local LERGs and HERGs with moderate radio powers typically have radio structures that are  unresolved on galaxy scales, it is important to note that they could possess extended structures on small (kpc to sub-kpc) physical scales. Indeed, the double-lobed structures found for some Seyfert galaxies typically manifest on scales of a few 100 pc \citep{uw84}, and some sources classified as FR0s in fact exhibit core-jet (`mini-FRI') morphologies on similar scales \citep[e.g.][]{baldi15}. In this context, high-resolution radio observations of HERGs with radio powers that are intermediate between those of high-power FRIIs on the one hand, and Seyfert galaxies on the other, could be particularly important for testing how strongly the black hole accretion mode is linked with the two traditional classifications of extended radio emission, i.e. HERGs with FRIIs and LERGs with FRIs, across all radio powers. We define these radio-intermediate HERGs\footnote{Note that we do not intend to imply that radio-intermediate HERGs are somehow intermediate between radio-quiet and radio-loud AGNs, as defined, for example, by optical-to-radio flux ratios, but rather that they are intermediate between high-radio-power FRIIs (log(L$_{\rm 1.4GHz}$) $>$ 25.0 W\,Hz$^{-1}$), which almost invariably have HERG spectra, and typical Seyfert galaxies (log(L$_{\rm 1.4GHz}$) $<$ 22.5 W\,Hz$^{-1}$).} to have radio powers in the range 22.5 $<$ log(L$_{\rm 1.4GHz}$) $<$ 25.0 W\,Hz$^{-1}$. Given that radio-intermediate HERGs show strong evidence for feedback, such observations could also be important for uncovering the scales on which the radio jets typically interact with the interstellar medium in their host galaxies, and hence help to improve models of galaxy evolution.

Despite their potential importance for understanding both accretion modes and jet-induced feedback, we lack knowledge of the detailed radio properties of radio-intermediate HERGs. Therefore, here we present high-resolution VLA radio observations of a sample of 16 local HERGs with intermediate radio powers ($z<0.1$; 23.0 $<$ log(L$_{\rm 1.4GHz}$) $< 24.0$ W\,Hz$^{-1}$), which were conducted to investigate their detailed radio morphologies down to sub-kpc physical scales. This is in order to test whether they are consistent with the double-lobed structures observed for high-power HERGs and some Seyfert galaxies, or rather the highly-compact or core-jet structures of LERG/FR0 sources. The radio spectra of these objects are also investigated. The paper is structured as follows. Details on the selection of the sample, the observations and the reduction, calibration and image production process are outlined in \S\ref{sec:sam_obs_red}. The results of the analysis of the images are presented in \S\ref{sec:analysis}, which are discussed and compared to results from the literature in \S\ref{sec:disc}. A summary of the study is presented in \S\ref{sec:sum}. Throughout this paper, we assume a cosmology described by $H_{0} = 73.0$ km\,s$^{-1}$\,Mpc$^{-1}$, $\Omega_{\rm m} = 0.27$ and $\Omega_{\rm \Lambda} = 0.73$, for consistency with our previous work.

\section{Sample selection, Observations and Reduction}
\label{sec:sam_obs_red}

\begin{table*}
	\centering
	\caption{Basic information for all 16 targets within the observed sample, selected as detailed in \S\protect\ref{sec:sample_sel}. Column 1 lists the full SDSS IDs for the targets with their in-text abbreviations in brackets, which match those of \citet{pierce19}. Column 2 lists the spectroscopic redshifts for the targets. Column 3 denotes the radio luminosities of the galaxies at 1.4\,GHz as derived from their NRAO VLA Sky Survey (NVSS) fluxes \protect\citep{con98}, or Faint Images of the Radio Sky at Twenty-cm (FIRST) flux \protect\citep{bec95} in the case of J1358; this was due to the contamination from an unassociated source lying within the angular scale of the NVSS beam. Columns 4, 5 and 6 provide the [OIII]\,$\rm \lambda5007$ luminosities, stellar masses and black hole masses for the targets, as derived from measurements provided by the MPA/JHU value-added catalogue for SDSS DR7\protect\footnotemark[2]. Column 7 lists the host galaxy types, as classified through visual inspection in \protect\citet{pierce19}.}
	\label{tab:target_info}
	\begin{tabular}{lcccccc}
		\hline
SDSS ID                           & z     & \makecell{log(L$_{\rm 1.4GHz}$)\\(W\,Hz$^{-1}$)} & \makecell{log(L$_{\rm O[III]}$)\\(W)} & log(M$_{*}$/M$_{\odot}$) & log(M$\rm_{BH}$/M$_{\odot}$) & Optical morph.   \\ \hline
SDSS J072528.47+434332.4 (J0725)  & 0.069 & 23.07                                                   & 33.90                         & 10.9                     & 8.3                          & Lenticular   \\
SDSS J075756.72+395936.1 (J0757)  & 0.066 & 23.98                                                   & 34.12                         & 10.8                     & 7.4                          & Merger           \\
SDSS J081040.29+481233.1 (J0810)  & 0.077 & 23.71                                                   & 33.54                         & 10.9                     & 7.9                          & Lenticular   \\
SDSS J083637.83+440109.5 (J0836)  & 0.055 & 23.94                                                   & 33.57                         & 10.3                     & 7.7                          & Merger           \\
SDSS J090239.51+521114.7 (J0902)  & 0.098 & 23.94                                                   & 33.25                         & 11.4                     & 8.5                          & Merger           \\
SDSS J095058.69+375758.8 (J0950)  & 0.041 & 23.37                                                   & 33.26                         & 10.8                     & 7.7                          & Edge-on disk \\
SDSS J110852.61+510225.6 (J1108)  & 0.070 & 23.07                                                   & 33.84                         & 11.0                     & 7.9                          & Lenticular   \\
SDSS J120640.31+104652.8 (J12064) & 0.089 & 23.41                                                   & 33.02                         & 11.0                     & 7.4                          & Spiral     \\
SDSS J123641.09+403225.7 (J1236)  & 0.096 & 23.08                                                   & 33.57                         & 10.5                     & 6.8                          & Edge-on disk \\
SDSS J124322.55+373858.0 (J1243)  & 0.086 & 23.36                                                   & 33.97                         & 11.2                     & 7.8                          & Merger           \\
SDSS J125739.56+510351.4 (J1257)  & 0.097 & 23.21                                                   & 33.33                         & 11.1                     & 7.5                          & Merger           \\
SDSS J132450.59+175815.0 (J1324)  & 0.086 & 23.36                                                   & 33.36                         & 10.7                     & 7.6                          & Spiral      \\
SDSS J135152.97+465026.1 (J1351)  & 0.096 & 23.26                                                   & 33.21                         & 10.6                     & 7.5                          & Merger           \\
SDSS J135817.73+171236.7 (J1358)  & 0.095 & 23.21                                                   & 33.93                         & 10.9                     & 7.9                          & Elliptical   \\
SDSS J141217.69+242735.9 (J1412)  & 0.069 & 23.20                                                   & 33.79                         & 11.1                     & 8.2                          & Elliptical  \\
SDSS J160953.46+133147.8 (J1609)  & 0.036 & 23.01                                                   & 32.86                         & 10.8                     & 7.5                          & Edge-on disk \\ \hline
	\end{tabular}
\end{table*}

\subsection{Sample selection}
\label{sec:sample_sel}

The sources studied in this work comprise a 94 per cent complete sample of 16 HERGs with radio powers in the range 23.0 $<$ log(L$_{\rm 1.4GHz}$) $<$ 24.0 W\,Hz$^{-1}$, low redshifts ($z <$ 0.1) and right ascension values ($\alpha$) in the range 07h 15m $< \alpha <$ 16h 45m, as selected from the catalogue of \citet[][]{bh12}. 
The complete sample fulfilling these criteria comprises 17 objects, but one object was not observed due to VLA scheduling constraints. Optical \textit{r}-band images for all the targets are presented in \citet[][]{pierce19}, whose sample also extended to lower radio powers (22.5 $<$ log(L$_{\rm 1.4GHz}$) $<$ 24.0); further details of the selection criteria are given in that paper. Note that we chose to concentrate on the higher end of the radio power range of the \citet[][]{pierce19} sample in order to optimise the signal-to-noise ratio achievable with relatively short on-source integration times (see \S\ref{sec:obs}), and hence improve the observing efficiency. We find that 15 out of the 16 observed sources (94 per cent) have unresolved radio structures at the limiting angular resolution of the FIRST survey ($\sim$5"). The VLA observations hence provided the opportunity to investigate the radio morphologies of the current sub-sample of 16 radio-intermediate HERGs on much smaller physical scales, to meet the goals of the project. 

In addition to the main sample, we observed a further two lower-power HERGs with the VLA (log(L$_{\rm 1.4GHz}$) $=$ 22.39 and 22.62) as part of this project. These were objects which initially fulfilled our radio power selection criterion based on low resolution NVSS data, which can be subject to source confusion, but were subsequently found to have lower radio powers based on higher resolution FIRST observations. Observations for these two objects are not included in the main body of the paper, but are presented in Appendix~\ref{sec:app}, for completeness.

The basic properties of the 16 targets in the main sample are summarised in Table~\ref{tab:target_info} -- the bracketed abbreviations of the target names listed here are used in the text throughout the paper, for simplicity and consistency with our previous work. The host types presented are visual classifications based on the morphological appearance of the galaxies at optical wavelengths, from \citet[][]{pierce19}. The sample contains a mixture of morphological types, with 31 per cent classified as early-type (elliptical or lenticular), 31 per cent as late-type (spiral or disk), and 38 per cent showing highly disturbed morphologies that are suggestive of galaxy mergers and interactions. The stellar masses of the galaxies lie in the range 10.3 $\leq$ log(M$_{*}$/M$_{\odot}$) $\leq$ 11.4, with a median of 10$^{10.9}$ M$_{\odot}$. The radio powers cover the range 23.01 $<$ log(L$_{\rm 1.4GHz}$) $<$ 23.98 W\,Hz$^{-1}$, but show a preference for lower radio powers within these limits and have a median value of 10$^{23.31}$ W\,Hz$^{-1}$.

\subsection{VLA observations}
\label{sec:obs}

\begin{table*}
\centering
\caption{Basic properties of the final images produced for all targets at all three frequencies. Beam sizes ($\theta_{\nu}$) and position angles ($\psi_{\nu}$) are provided, with the latter measured with the positive direction from North to East on the sky. Root-mean-square (rms; $\sigma_{\nu}$) values, as measured from the residual background variations in boxes placed in suitable regions close to the targets, are also presented. The dates of the target observations are also listed in the final column.}
\label{tab:image_prop}
\begin{tabular}{ccccccccccc}
\hline
Name   & \makecell{$\rm \theta_{1.5}$\\(arcsec)} & \makecell{$\rm \psi_{1.5}$\\(deg)} & \makecell{$\rm \sigma_{1.5}$\\($\mu$Jy)} & \makecell{$\rm \theta_{4.5}$\\(arcsec)} & \makecell{$\rm \psi_{4.5}$\\(deg)} & \makecell{$\rm \sigma_{4.5}$\\($\mu$Jy)} & \makecell{$\rm \theta_{7.5}$\\(arcsec)} & \makecell{$\rm \psi_{7.5}$\\(deg)} & \makecell{$\rm \sigma_{7.5}$\\($\mu$Jy)} & Obs. date  \\ \hline
J0725  & 1.76 $\times$ 1.40                                                                     & 81.2                                                                              & 86.0                                                                                    & 0.71 $\times$ 0.50                                                                     & 85.1                                                                                  & 16.4                                                                                    & 0.41 $\times$ 0.28                                                                     & -82.5                                                                             & 12.0                                                                                    & 2018-03-25 \\
J0757  & 1.98 $\times$ 1.40                                                                     & -84.4                                                                             & 58.4                                                                                    & 0.74 $\times$ 0.54                                                                     & 78.7                                                                                  & 20.6                                                                                    & 0.42 $\times$ 0.29                                                                     & -88.0                                                                             & 21.9                                                                                    & 2018-03-25 \\
J0810  & 1.66 $\times$ 1.19                                                                     & 89.2                                                                              & 117.0                                                                                   & 0.77 $\times$ 0.51                                                                     & 73.1                                                                                  & 18.1                                                                                    & 0.42 $\times$ 0.28                                                                     & 83.4                                                                              & 16.6                                                                                    & 2018-03-25 \\
J0836  & 1.80 $\times$ 1.28                                                                     & -81.2                                                                             & 66.5                                                                                    & 0.70 $\times$ 0.50                                                                     & 85.0                                                                                  & 26.8                                                                                    & 0.40 $\times$ 0.28                                                                     & -84.7                                                                             & 17.9                                                                                    & 2018-03-25 \\
J0902  & 1.62 $\times$ 1.24                                                                     & 87.7                                                                              & 71.2                                                                                    & 0.88 $\times$ 0.50                                                                     & 76.0                                                                                  & 18.2                                                                                    & 0.45 $\times$ 0.28                                                                     & 86.8                                                                              & 23.8                                                                                    & 2018-03-28 \\
J0950  & 1.70 $\times$ 1.35                                                                     & 37.7                                                                              & 90.6                                                                                    & 0.78 $\times$ 0.45                                                                     & -84.1                                                                                 & 17.0                                                                                    & 0.45 $\times$ 0.28                                                                     & -84.3                                                                             & 16.2                                                                                    & 2018-03-28 \\
J1108  & 2.01 $\times$ 1.07                                                                     & 82.5                                                                              & 77.2                                                                                    & 0.60 $\times$ 0.45                                                                     & 76.4                                                                                  & 14.0                                                                                    & 0.52 $\times$ 0.35                                                                     & 28.1                                                                              & 16.1                                                                                    & 2018-04-14 \\
J12064 & 2.26 $\times$ 1.58                                                                     & -36.8                                                                             & 182.4                                                                                   & 0.83 $\times$ 0.45                                                                     & -52.8                                                                                 & 17.5                                                                                    & 0.51 $\times$ 0.30                                                                     & -53.7                                                                             & 15.3                                                                                    & 2018-04-14 \\
J1236  & 1.63 $\times$ 1.16                                                                     & -85.4                                                                             & 58.3                                                                                    & 0.67 $\times$ 0.45                                                                     & -83.0                                                                                 & 19.2                                                                                    & 0.42 $\times$ 0.29                                                                     & -84.4                                                                             & 13.8                                                                                    & 2018-04-14 \\
J1243  & 2.74 $\times$ 1.47                                                                     & 78.3                                                                              & 280.0                                                                                   & 0.66 $\times$ 0.49                                                                     & 88.2                                                                                  & 18.0                                                                                    & 0.41 $\times$ 0.29                                                                     & -85.0                                                                             & 13.8                                                                                    & 2018-04-14 \\
J1257  & 1.54 $\times$ 1.22                                                                     & 69.1                                                                              & 50.8                                                                                    & 0.57 $\times$ 0.45                                                                     & 45.5                                                                                  & 13.2                                                                                    & 0.34 $\times$ 0.27                                                                     & 45.5                                                                              & 13.0                                                                                    & 2018-04-12 \\
J1324  & 1.48 $\times$ 1.20                                                                     & -48.8                                                                             & 81.5                                                                                    & 0.55 $\times$ 0.47                                                                     & -46.1                                                                                 & 12.3                                                                                    & 0.33 $\times$ 0.28                                                                     & -45.6                                                                             & 11.5                                                                                    & 2018-04-12 \\
J1351  & 1.58 $\times$ 1.27                                                                     & 0.9                                                                               & 57.2                                                                                    & 0.57 $\times$ 0.47                                                                     & 29.3                                                                                  & 13.6                                                                                    & 0.37 $\times$ 0.28                                                                     & 28.1                                                                              & 12.9                                                                                    & 2018-04-12 \\
J1358  & 1.58 $\times$ 1.30                                                                     & -86.8                                                                             & 101.9                                                                                   & 0.58 $\times$ 0.47                                                                     & -50.7                                                                                 & 12.6                                                                                    & 0.35 $\times$ 0.29                                                                     & -53.5                                                                             & 15.9                                                                                    & 2018-04-12 \\
J1412  & 1.43 $\times$ 1.27                                                                     & 42.2                                                                              & 60.4                                                                                    & 0.52 $\times$ 0.47                                                                     & 7.6                                                                                   & 21.8                                                                                    & 0.31 $\times$ 0.28                                                                     & 5.8                                                                               & 14.3                                                                                    & 2018-05-17 \\
J1609  & 1.38 $\times$ 1.15                                                                     & 6.8                                                                               & 84.9                                                                                    & 0.52 $\times$ 0.45                                                                     & 0.6                                                                                   & 25.6                                                                                    & 0.31 $\times$ 0.27                                                                     & 0.6                                                                               & 13.8                                                                                    & 2018-05-17\\ \hline               
\end{tabular}
\end{table*}

Continuum L- and C-band observations of the 16 sources in the sample were obtained using the Karl G. Jansky Very Large Array (VLA) in the A-array configuration. These were taken from 2018 March 25 to 2018 May 17. The use of these two frequency bands allowed for the detection of both diffuse (L-band) and more compact (C-band) emission structures, whilst also opening up the possibility of determining spectral indices for the sources in the frequency range covered. The L-band observations were conducted using the default configuration of two 8-bit samplers each with 64 $\times$ 1.0 MHz-width channels, yielding a total bandwidth of 1024 MHz centred on a frequency of 1.5 GHz. Two 8-bit samplers were also used for the C-band observations, but were composed of 64 $\times$ 2.0 MHz-width channels (i.e. each with total width 1024 MHz) and centred on frequencies of 4.5 GHz and 7.5 GHz. This allowed the C-band to be separated into two distinct sub-bands, which could be considered separately during subsequent analysis and image production. 

\footnotetext[2]{Available at: \\ https://wwwmpa.mpa-garching.mpg.de/SDSS/DR7/.}

On-source integration times for L- and C-band observations were $\sim$5--6 minutes and $\sim$10--12 minutes, respectively. One of the two standard flux calibrators 3C 147 or 3C 286 was observed at the beginning of each set of observations in both L- and C-bands, and these were used for flux and bandpass calibration of the target observations in both bands. The on-source times in the L-band were $\sim$6--7 minutes and $\sim$8--10 minutes for 3C 147 and 3C 286, respectively, with $\sim$2 minute on-source times being used for both calibrators in the C-band. Suitable phase calibrators, selected from the list of standard VLA calibrators\footnote[3]{Available at: \\ https://science.nrao.edu/facilities/vla/observing/callist.}, were observed before and after each target observation for $\sim$1--3.5 minutes and $\sim$2--4 minutes in the L- and C-bands, respectively. These were used to determine calibration solutions for the complex antenna gains. 

All data reduction, calibration and image production was performed using standard procedures in the \textit{Common Astronomy Software Applications} (CASA, v.~5.4.1) package, specifically designed for the post-processing of radio-astronomical data \citep[see][]{casa}. Manual flagging of spurious bad data caused by issues related to individual antenna problems, bad baseline correlations or radio frequency interference was carried out on all calibrator targets prior to calibration of the science target data. Such flagging was also performed on the science target data after the relevant calibration procedures had been applied. This resulted in significant data loss in some cases; we estimate that up to $\sim$50 per cent of the data was lost for the L-band target observations, in particular, but only $\sim$10--20 per cent for the C-band observations.

Following this, self-calibration of the science target data was also performed iteratively, firstly using phase-only solutions to remove any residual phase errors that remained after the primary calibration procedures. These were then followed by joint amplitude-phase solutions to correct for residual amplitude errors. This was done until the quality of the subsequent image was not further improved; in some cases, the initial self-calibration provided no improvements, either due to the faintness of the target or to bright sources in the surrounding field affecting the solutions. 

All continuum images were produced using the CASA task \textsc{clean}. Images were made for the observations of each target in the L-band and in both of the two sub-bands of the C-band, to provide continuum images centred on frequencies of 1.5 GHz, 4.5 GHz and 7.5 GHz. Pixel sizes of 0.4 $\times$ 0.4 arcseconds and 0.1 $\times$ 0.1 arcseconds were used for the L-band and C-band images, respectively, and natural weighting was used in all cases. The limiting angular resolution achieved in the L-band images for the main sample, estimated by averaging the two beam axes in each case, covered the range 1.27 -- 2.11 arcseconds, with a median of 1.44 arcseconds. Derived in the same way, the values for the C-band were found to cover the range 0.49 -- 0.69 arcseconds (median $=$ 0.57 arcseconds) in the images at 4.5 GHz, and from 0.29 -- 0.44 arcseconds (median $=$ 0.34 arcseconds) in the images at 7.5 GHz. The median root-mean-square (rms) values for residual background variations in the final images, measured in boxes placed in suitable regions close to the targets, are 79.4 $\mu$Jy\,beam$^{-1}$, 17.8 $\mu$Jy\,beam$^{-1}$ and 14.8 $\mu$Jy\,beam$^{-1}$ for the images at 1.5 GHz, 4.5 GHz and 7.5 GHz, respectively. The restoring beam sizes and rms values for all of the final continuum images at all frequencies are listed in Table~\ref{tab:image_prop}, along with dates of the individual target observations.

\section{Image analysis and Results}
\label{sec:analysis}

\begin{figure*}
    \centering
    \includegraphics[width=0.9\textwidth]{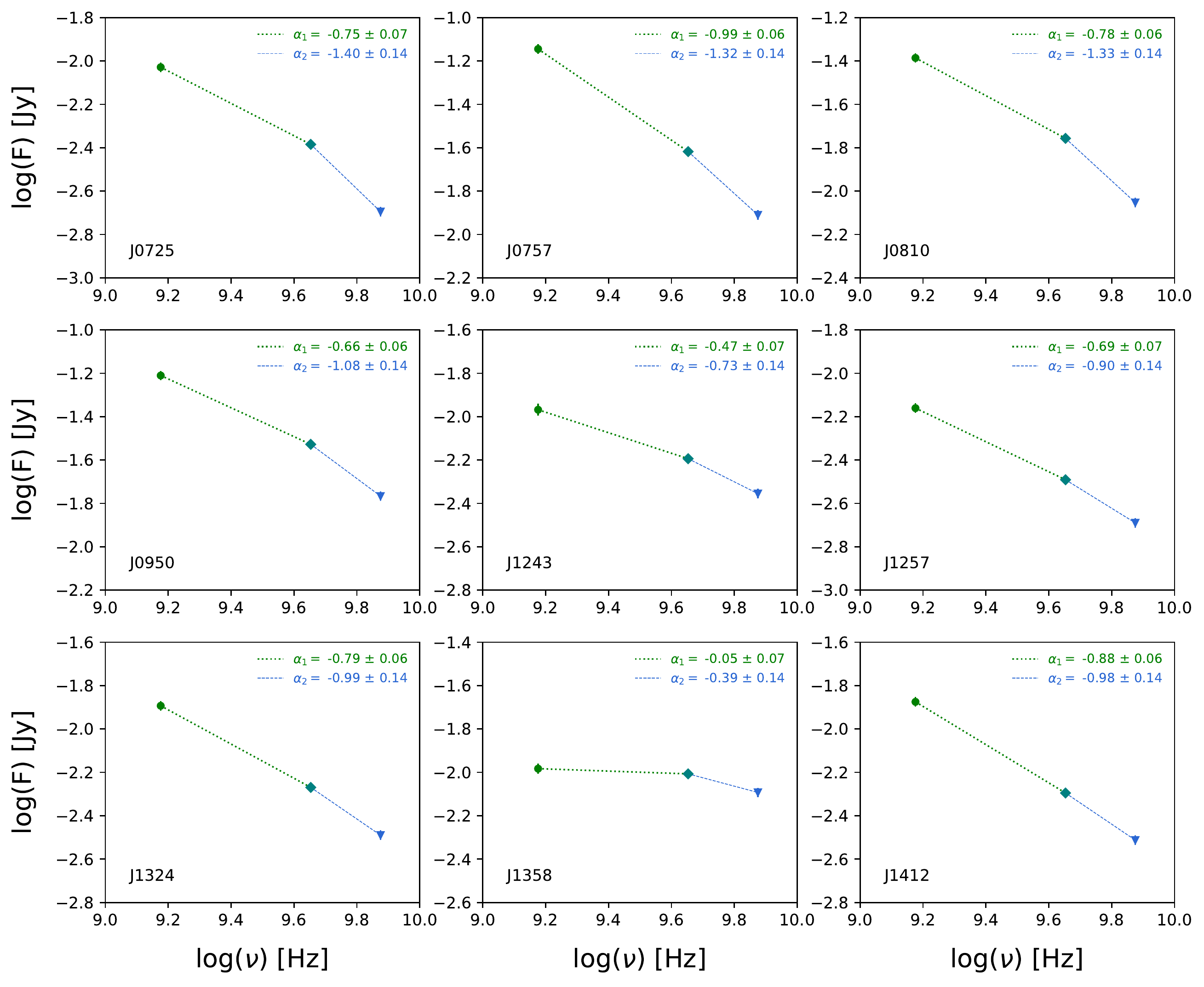}
    \caption{Basic radio spectra for the 9 targets that have unresolved structures in the images at all three frequencies, and hence appear as point sources. The flux measurements were determined using the CASA task \textsc{imfit}. The uncertainties associated with these measurements are plotted, but are too small to be visible on the graphs. Estimations of the spectral indices between 1.5 GHz and 4.5 GHz ($\alpha_1$) and between 4.5 GHz and 7.5 GHz ($\alpha_2$) are indicated on each graph alongside their associated errors $-$ these are also listed in Table~\ref{tab:ps_fluxes}. These were derived from the measured fluxes at the limits of the aforementioned ranges, as appropriate. The target names are also indicated in the bottom-left corner of the graphs.}
    \label{fig:3pt_spec}
\end{figure*}

The final continuum images produced for all 16 radio-intermediate HERGs at each of the three frequencies (1.5 GHz, 4.5 GHz and 7.5 GHz) were used to investigate the detailed morphological structure of the radio emission, to meet the main objectives of the project. We find that there is a mixture of unresolved and extended sources within the sample, which are discussed separately below (in \S\ref{sec:point} and \S\ref{sec:ext}, respectively). The flux and scale of the emission at the three frequencies was determined from all of the target images, derived as described below and presented in Tables~\ref{tab:ps_fluxes} and \ref{tab:ext_info}. We also present radio spectra for the unresolved sources, which are shown in Figure~\ref{fig:3pt_spec}. Contour maps of the extended sources at each frequency are provided in Figure~\ref{fig:cont_a}.

\subsection{Unresolved sources}
\label{sec:point}

\begin{figure}
    \centering
    \includegraphics[width=0.9\columnwidth]{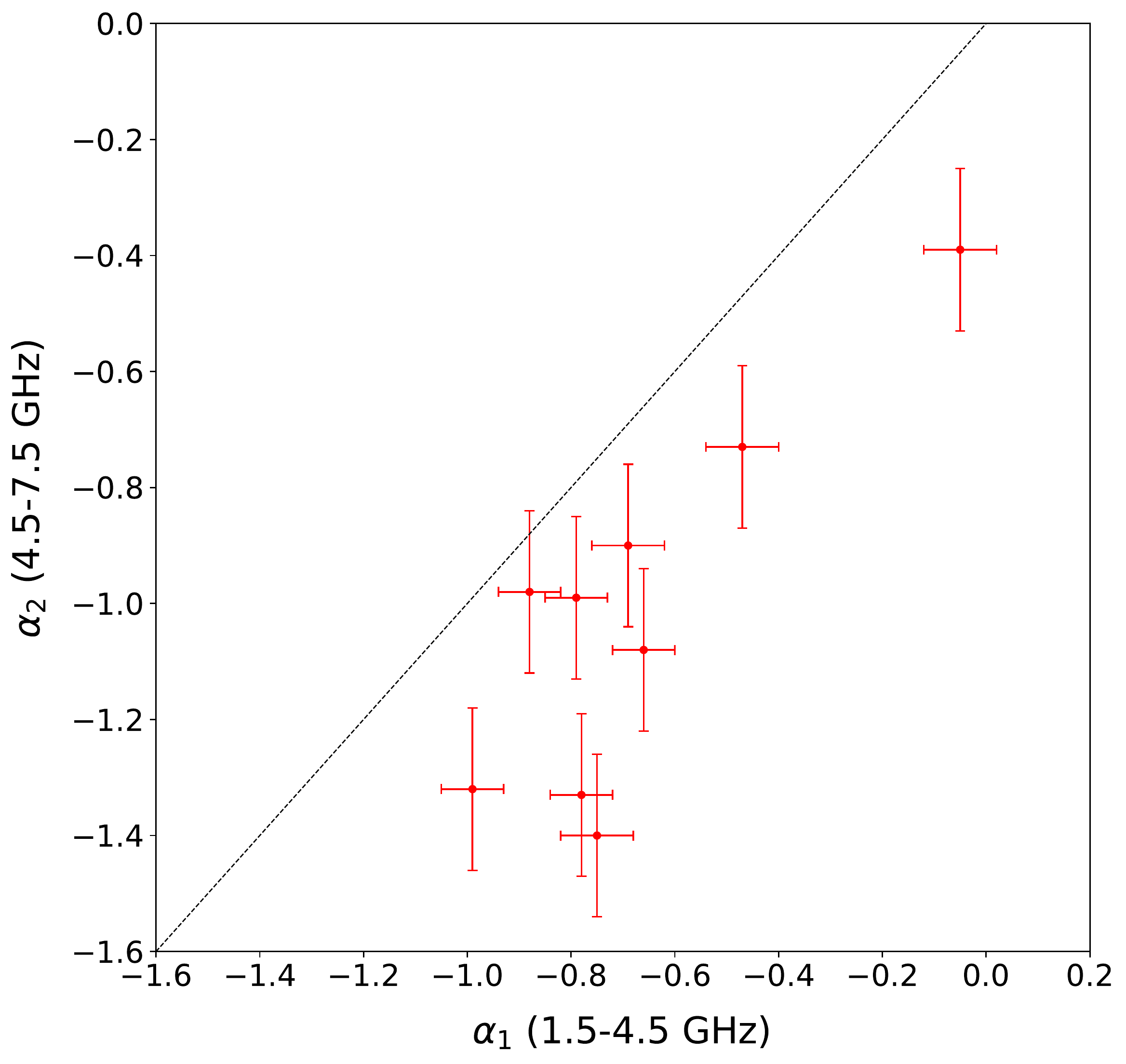}
    \caption{The spectral indices from 4.5 GHz to 7.5 GHz ($\alpha_2$) against those from 1.5 GHz to 4.5 GHz ($\alpha_1$) for the unresolved sources in the sample. All points are seen to lie below the line of one-to-one correspondence (black), highlighting the systematic spectral steepening that occurs at higher frequencies. The errors on the points are those listed in Table~\ref{tab:ps_fluxes}.}
    \label{fig:spec_steepening}
\end{figure}

\begin{table*}
\centering
\caption{Flux measurements and spectral indices for all 9 sources in the sample with unresolved radio structures in the images at all three frequencies, whose basic radio spectra are presented in Figure~\ref{fig:3pt_spec}. The peak flux values at each frequency ($f_{\nu}^{P}$) are provided by the double-Gaussian fits of the target emission performed by the CASA task \textsc{imfit}. Separate spectral indices for the power law slopes between 1.5 GHz and 4.5 GHz ($\alpha_1$) and between 4.5 GHz and 7.5 GHz ($\alpha_2$) are also listed. The maximum angular extents of the structures are provided in the penultimate column (in arcseconds), as estimated using the average of the two beam axes for the images at 7.5 GHz. Estimated physical sizes derived from these values are listed in parsecs in the final column.}
\label{tab:ps_fluxes}
\begin{tabular}{cccccccccc}
\hline
Name  & \makecell{F$_{\rm NVSS}$\\(mJy)} & \makecell{F$_{\rm FIRST}$\\(mJy)} & \makecell{$f^{P}_{\rm 1.5}$\\(mJy)} & \makecell{$f^{P}_{\rm 4.5}$\\(mJy)} & \makecell{$f^{P}_{7.5}$\\(mJy)}    & \makecell{$\alpha_{1}$\\(1.5-4.5 GHz)}       & \makecell{$\alpha_{2}$\\(4.5-7.5 GHz)}     & \makecell{$\phi^{\rm lim}$\\(arcsec)} & \makecell{$r_{\rm lim}$\\(pc)} \\ \hline
J0725 & 11.0           & 10.8            & 9.36 $\pm$ 0.48   & 4.12 $\pm$ 0.21   & 2.02 $\pm$ 0.10  & -0.746 $\pm$ 0.066 & -1.40 $\pm$ 0.14 & 0.35             & 440           \\
J0757 & 99.7           & 99.2            & 71.6 $\pm$ 3.6    & 24.1 $\pm$ 1.2    & 12.30 $\pm$ 0.62 & -0.991 $\pm$ 0.064 & -1.32 $\pm$ 0.14 & 0.36             & 430           \\
J0810 & 37.8           & 49.1            & 41.1 $\pm$ 2.1    & 17.50 $\pm$ 0.88  & 8.85 $\pm$ 0.44  & -0.777 $\pm$ 0.064 & -1.33 $\pm$ 0.14 & 0.35             & 490           \\
J0950 & 66.6           & 68.3            & 61.6 $\pm$ 3.1    & 29.7 $\pm$ 1.5    & 17.12 $\pm$ 0.86 & -0.665 $\pm$ 0.064 & -1.08 $\pm$ 0.14 & 0.37             & 280           \\
J1243 & 13.6           & 12.6            & 10.76 $\pm$ 0.69  & 6.40 $\pm$ 0.32   & 4.42 $\pm$ 0.22  & -0.473 $\pm$ 0.074 & -0.73 $\pm$ 0.14 & 0.35             & 540           \\
J1257 & 7.5            & 7.2             & 6.91 $\pm$ 0.35   & 3.23 $\pm$ 0.16   & 2.04 $\pm$ 0.10  & -0.692 $\pm$ 0.065 & -0.90 $\pm$ 0.14 & 0.31             & 520           \\
J1324 & 13.6           & 14.5            & 12.79 $\pm$ 0.65  & 5.38 $\pm$ 0.27   & 3.24 $\pm$ 0.16  & -0.789 $\pm$ 0.065 & -0.99 $\pm$ 0.14 & 0.31             & 470           \\
J1358 & 17.8           & 7.9             & 10.40 $\pm$ 0.54  & 9.84 $\pm$ 0.49   & 8.08 $\pm$ 0.40  & -0.050 $\pm$ 0.066 & -0.39 $\pm$ 0.14 & 0.32             & 540           \\
J1412 & 14.9           & 13.2            & 13.34 $\pm$ 0.67  & 5.07 $\pm$ 0.26   & 3.07 $\pm$ 0.16  & -0.881 $\pm$ 0.065 & -0.98 $\pm$ 0.14 & 0.30             & 370          
\\ \hline
\end{tabular}
\end{table*}

\begin{figure}
    \centering
    \includegraphics[width=0.95\columnwidth]{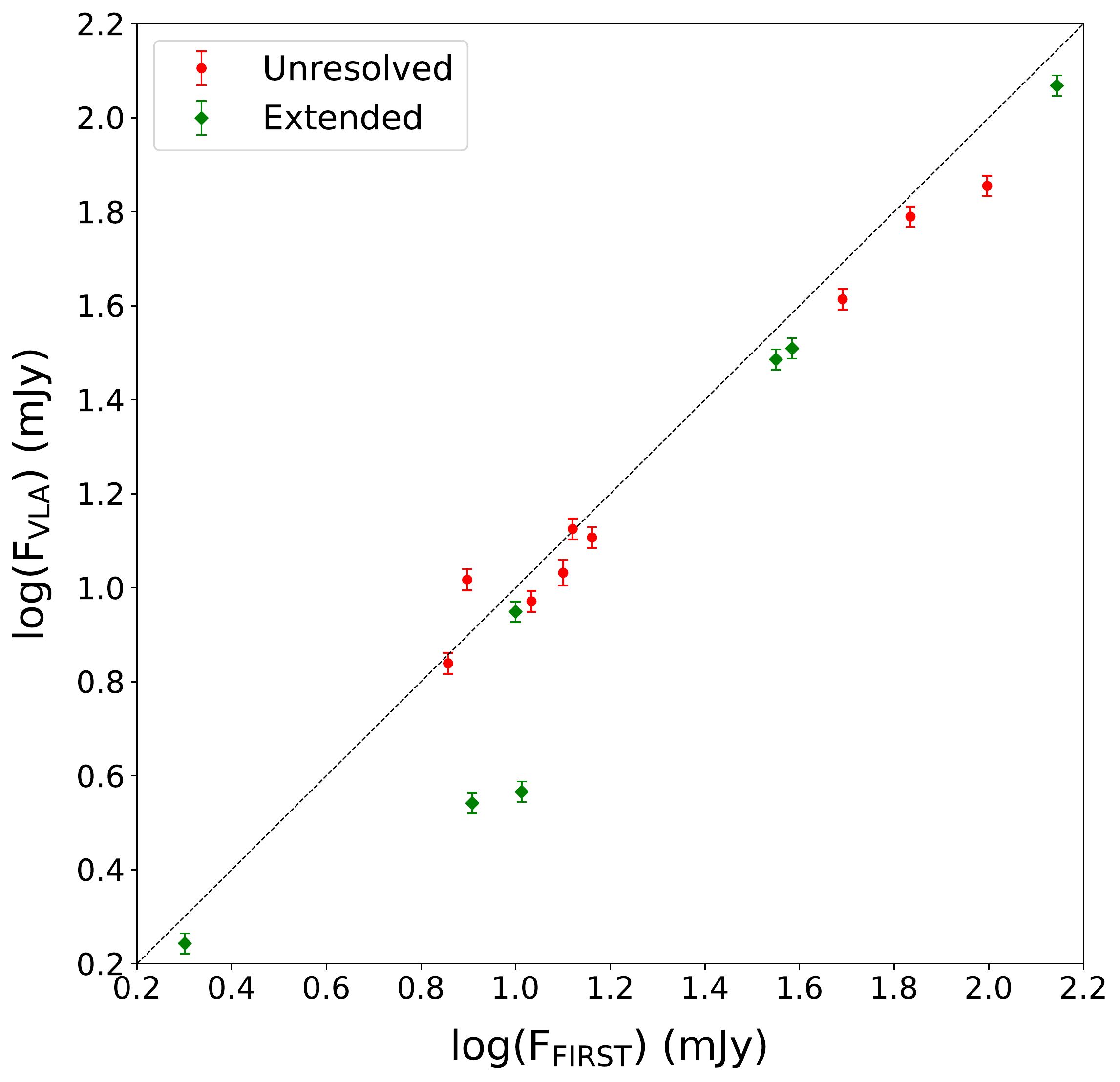}
    \caption{The fluxes measured from the VLA images at 1.5 GHz for all unresolved (red points) and extended (green points) sources in the sample, plotted against their FIRST fluxes. Errors are as listed in Tables~\ref{tab:ps_fluxes} and \ref{tab:ext_info}.}
    \label{fig:FIRSTvsVLA}
\end{figure}

The first step in characterising the morphology of the radio emission involved separating the sample into resolved and unresolved structures. In order to evaluate whether the radio structures were resolved or unresolved at the given image resolution, model Gaussian profiles were fitted to the target emission using the CASA task \textsc{imfit}. The measured angular full-width half maximum (FWHM), position angle and peak flux provided by the fits were then compared with the restoring beam parameters and total flux measured for each target, to determine whether the emission was more characteristic of an unresolved or extended source. 

From this, it is seen that 9 of the 16 galaxies in the sample (56 per cent) have unresolved structures in the VLA observations at all three frequencies, meaning that they are compact to an angular resolution of $\sim$0.3 arcseconds. In terms of projected physical scale, this is equivalent to $\sim$220--500 pc for these galaxies (median $=$ 420 pc). The limiting angular sizes and corresponding maximum physical diameters of the emission for each of the unresolved sources are reported in Table~\ref{tab:ps_fluxes}, as obtained from the \textsc{imfit} fits to the images at 7.5 GHz. 

Since the radio structures for these compact sources are unresolved in the images at all three frequencies, we considered the peak flux values determined by the \textsc{imfit} task to be reasonable estimates for the total flux values for each target at each frequency. Errors in the peak flux measurements were estimated by taking into account contributions from the fitting errors provided by \textsc{imfit}, the background noise in the image \citep[following][]{kle03}, and the uncertainty in the flux calibration models \citep[considering the upper limit of 5 per cent determined by][]{per17}. Based on the assumption that the sources are point-like at all frequencies, it was then possible to produce basic radio spectra for the targets and estimate the slope of the power law (for $\rm F_{\nu} \propto \nu^{\alpha}$) between 1.5 GHz and 4.5 GHz ($\alpha_1$) and between 4.5 GHz and 7.5 GHz ($\alpha_2$). These spectra are presented in Figure~\ref{fig:3pt_spec}, and the peak flux values and estimated power law slopes are also listed in Table~\ref{tab:ps_fluxes}. 

From the spectra, it is seen that the majority (7 out of 9) of the unresolved sources show steep power laws between 1.5 GHz and 4.5 GHz ($-0.99 < \alpha_1 < -0.67$), which are characteristic of standard predictions for optically-thin synchrotron emission. The objects J1243 and J1358, however, have flatter spectra between these frequencies ($\alpha_1 = -0.47$ and -0.05), and J1358 remains relatively flat between 4.5 GHz and 7.5 GHz ($\alpha_2= -0.39$). All of the unresolved objects show steeper spectra between 4.5 GHz and 7.5 GHz (median $\alpha_2=$ -0.99), regardless of their spectral index between 1.5 GHz and 4.5 GHz (median $\alpha_1=$ -0.75), with a median difference in spectral index of $\Delta\alpha=0.33$ seen between the two ranges. This is shown clearly in Figure~\ref{fig:spec_steepening}, where we plot $\alpha_2$ against $\alpha_1$ for the unresolved sources and it is seen that all points lie below the line of equality between the two.

Although we find that these sources are unresolved at the resolution of our VLA observations, it is important to assess whether the sources are truly compact in nature, or rather have extended emission on larger scales that has been resolved out. To investigate this, we here compare the 1.5 GHz peak fluxes derived from the VLA observations (here F$\rm_{VLA}$) with their 1.4 GHz FIRST fluxes (F$_{\rm FIRST}$); these are plotted against each other in Figure~\ref{fig:FIRSTvsVLA}, for both the unresolved and extended sources (see \S\ref{sec:ext} for discussion of the latter). Here, we see that the majority of the unresolved sources lie close to the line of equality between the two, and in fact 6 out of 9 have F$_{\rm FIRST}$ values that lie within 3$\sigma$ of their F$\rm_{VLA}$ values, suggesting that in most cases the sources have little diffuse emission and are genuinely compact on sub-kpc scales. 

However, there is still a tendency for the VLA 1.5 GHz fluxes to lie below the FIRST values, with a median F$\rm_{VLA}$/F$\rm_{FIRST}$ ratio of 0.88. This could at least partly be explained by the slight difference in the effective frequencies of the two observations: using the median spectral index of the unresolved objects between 1.5 and 4.5 GHz, $\alpha_1=$ -0.75, the expected ratio of the fluxes at 1.5 and 1.4 GHz would be F$\rm_{VLA}$/F$\rm_{FIRST}=0.95$. Furthermore, when flagging the L-band data, we found that more was lost towards the lower frequencies within the band, which would shift the effective frequency to higher values and further lower the expected F$\rm_{VLA}$/F$\rm_{FIRST}$ ratio; if the effective central frequency was shifted by an additional 0.1 GHz to 1.6 GHz, for example, the expected value would be F$\rm_{VLA}$/F$\rm_{FIRST}=0.90$.

However, two of the sources that are unresolved in our VLA observations, J0757 and J0810, have F$\rm_{VLA}$ values that are much lower than their FIRST values (differences of 7.7$\sigma$ and 3.9$\sigma$, respectively). Despite the fact that their emission structures appear to be unresolved in both the FIRST and VLA images, there is still the possibility that the lower fluxes in these cases could be explained by the loss of detection of diffuse emission at the higher angular resolutions achieved. 

On the other hand, the final unresolved source (J1358) has a VLA 1.5 GHz flux that lies 4.6$\sigma$ \textit{above} its FIRST value, and the discrepancy hence cannot be explained in the same way. The flat spectrum observed for J1358, however, could suggest that it is a beamed, variable source, which could explain the difference in this case. Further discussion on the unresolved sources is reserved for \S\ref{sec:disc}.

\subsection{Extended sources}
\label{sec:ext}

All of the 7 remaining galaxies in the sample (44 per cent) show at least marginally resolved structures in the images at all three frequencies, i.e. even at the typical angular resolution of the L-band images ($\sim$1.4 arcseconds). Contour maps at all three frequencies are shown for these galaxies in Figure~\ref{fig:cont_a}, and their contour levels and beam parameters are presented in Table~\ref{tab:ext_info}. A brief description of the emission for each of the extended sources follows here, including comparisons with the optical appearance of the host galaxies in the \textit{r}-band images of \citet[][]{pierce19}.

\textbf{J0836}. This source is clearly extended in the images at all three observing frequencies. It exhibits a tail of emission that propagates to $\sim$4 kpc in projected radial extent from its centre, which is most clearly visible in the two C-band images. The C-band images also reveal a potential double-lobed structure that appears to be symmetric about the optical centre of the galaxy (particularly at 7.5 GHz), with projected diameter $\sim$2 kpc. The emission is confined to the high-surface-brightness structure of the galaxy that is visible at optical wavelengths.

\textbf{J0902}. This source displays emission that is highly collimated on small scales, but shows a `two-tailed', `S-shape' emission structure at all three frequencies on larger scales. The emission in the outer regions of the tails appears to become diffuse below the detection level when moving towards higher frequencies, due to the increasing resolution. The emission extends over a large proportion of the optical extent of the host galaxy, covering around 16 kpc in maximum projected diameter, with the ends of the tails almost propagating into the optical galaxy halo. Interestingly, its appearance is reminiscent of the innermost kiloparsec of the synthetic radio map produced by \citet{murthy19} to explain the features of the low-luminosity radio source B2 0258$+$35, made by simulating the effects of an inclined jet interacting with an edge-on disk of dense gas \citep[following][]{muk18a,muk18b}. Although we do not see any indications of such a disk at optical wavelengths, this host galaxy appears to have undergone a merger \citep[see][]{pierce19}, and so it is possible that a higher density of gas could have been supplied to its nuclear regions by this event. Similar structures have also been observed in nearby Seyfert galaxies, with jet precession and/or disruption being the favoured explanations in this instance \citep{kharb06,kharb10}. Detailed study of optical to infrared emission spectra in these regions could be used to further investigate this possibility, to search for signs of kinematic disturbance in the emission lines that would indicate a strong jet-cloud interaction.

\begin{figure*}
    \centering
    \includegraphics[width=\textwidth]{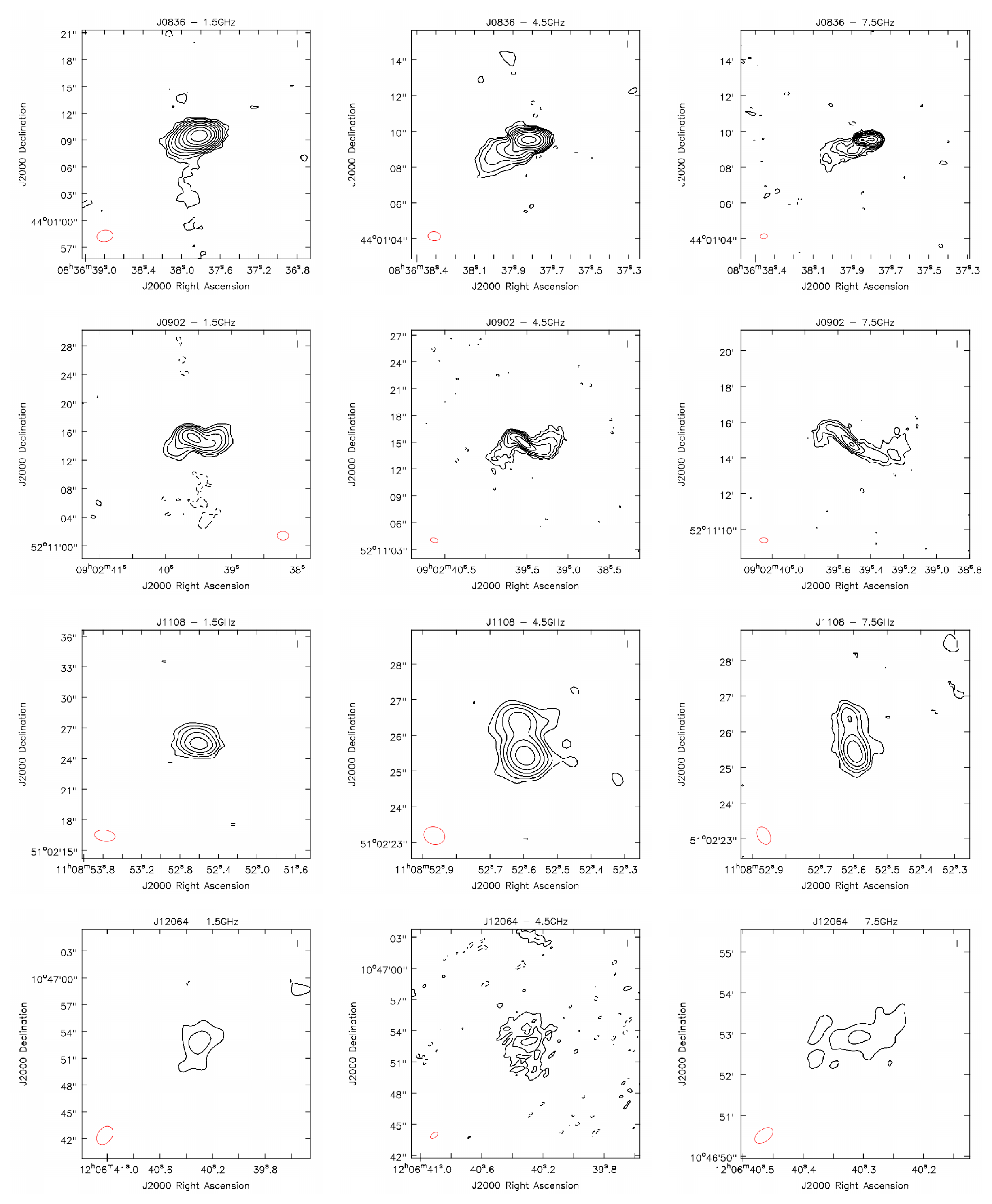}
    \caption{Contour plots for the first four targets in the sample that exhibit extended radio emission at 1.5 GHz (left column), 4.5 GHz (middle column) and 7.5 GHz (right column). Positive contours (solid lines) begin at $3\sigma$ and typically increase by subsequent factors of two; exact contour levels are listed in Table~\ref{tab:ext_info}. Negative contours at the $-3\sigma$ level are indicated with dashed lines. The red ellipses indicate the approximate beam shape for each image. The images are oriented such that North is up and East is left. The ranges of right ascension and declination covered are indicated in each case.}
    \label{fig:cont_a}
\end{figure*}

\begin{figure*}
    \centering
    \includegraphics[width=\textwidth]{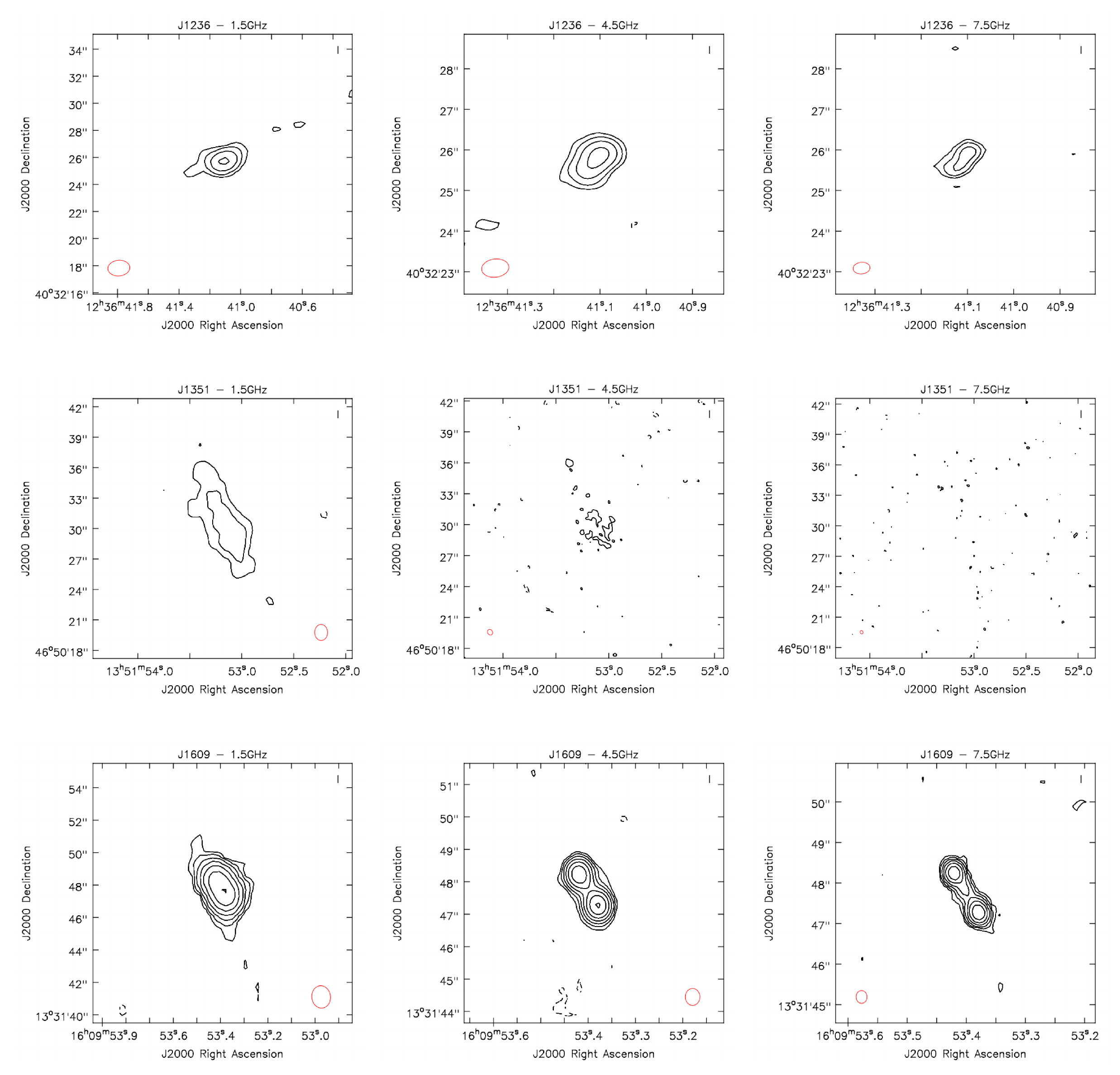}
    \contcaption{}
    \label{fig:cont_b}
\end{figure*}

\begin{table*}
\centering
\caption{Measurements of the peak fluxes ($f_{\nu}^{p}$), total fluxes ($f_{\nu}^{T}$), maximum angular sizes ($\phi_{\nu}^{\rm max}$) and projected physical diameters ($d_{\nu}^{\rm max}$) of the radio emission for the extended sources in the sample. Contour levels for the continuum maps displayed in Figure~\ref{fig:cont_a} are also presented, as multiples of the background rms measurements in the images ($\rm \sigma_{rms}$; $\mu$Jy, final column). The NVSS \citep{con98} and FIRST \citep{bec95} fluxes, measured at 1.4 GHz, are also presented for comparison. All flux measurements are given in mJy. The maximum angular sizes of the structures are listed in arcseconds and their corresponding maximum physical extents in kpc. The three rows for each target indicate the values for the 1.5 GHz, 4.5 GHz and 7.5 GHz maps, respectively.}
\label{tab:ext_info}
\begin{tabular}{ccccccccc}
\hline
Name   & \makecell{F$_{\rm NVSS}$\\(mJy)} & \makecell{F$_{\rm FIRST}$\\(mJy)} & \makecell{$f_{\rm \nu}^{P}$\\(mJy)} & \makecell{$f_{\rm \nu}^{T}$\\(mJy)} & \makecell{$\phi_{\nu}^{\rm max}$\\(arcsec)} & \makecell{$d_{\nu}^{\rm max}$\\(kpc)} & \makecell{Contour levels\\($n\sigma_{\rm rms}$)} & \makecell{$\sigma_{\rm rms}$\\($\mu$Jy)} \\ \hline
J0836  & 130.8          & 139.3           & 93.6 $\pm$ 4.7    & 117.0 $\pm$ 5.9   & 7.0                    & 7.2                 & {[}-3, 3, 6, 12, 24, 48, 96, 192, 384, 768{]} & 66.5                         \\
       &                &                 & 29.3 $\pm$ 1.5    & 55.7 $\pm$ 2.8    & 4.5                    & 4.7                 & {[}-3, 3, 6, 12, 24, 48, 96, 192, 384, 768{]} & 26.8                         \\
       &                &                 & 15.18 $\pm$ 0.76  & 35.3 $\pm$ 1.8    & 3.6                    & 3.7                 & {[}-3, 3, 6, 12, 24, 48, 96, 192, 384, 768{]} & 17.9                         \\
J0902  & 39.5           & 38.4            & 9.96 $\pm$ 0.50   & 32.3 $\pm$ 1.6    & 9.4                    & 16.4                & {[}-3, 3, 6, 12, 24, 48, 96{]}                & 71.2                         \\
       &                &                 & 3.15 $\pm$ 0.16   & 16.47 $\pm$ 0.83  & 8.7                    & 15.2                & {[}-3, 3, 6, 12, 24, 48, 96, 192, 384, 768{]} & 18.2                         \\
       &                &                 & 1.386 $\pm$ 0.073 & 8.32 $\pm$ 0.44   & 5.1                    & 8.9                 & {[}-3, 3, 6, 12, 24, 48, 96, 192, 384, 768{]} & 23.8                         \\
J1108  & 10.9           & 10.0            & 6.33 $\pm$ 0.33   & 8.89 $\pm$ 0.47   & 4.8                    & 6.1                 & {[}-3, 3, 6, 12, 24, 48{]}                    & 77.2                         \\
       &                &                 & 2.28 $\pm$ 0.11   & 2.07 $\pm$ 0.11   & 2.2                    & 2.8                 & {[}-3, 3, 6, 12, 24, 48, 96{]}                & 14.0                         \\
       &                &                 & 1.533 $\pm$ 0.078 & 2.58 $\pm$ 0.14   & 2.0                    & 2.6                 & {[}-3, 3, 6, 12, 24, 48{]}                    & 16.1                         \\
J12064 & 14.2           & 10.3            & 2.04 $\pm$ 0.21   & 3.68 $\pm$ 0.32   & 3.8                    & 6.1                 & {[}-3, 3, 6{]}                                & 182.4                        \\
       &                &                 & 0.321 $\pm$ 0.024 & 2.16 $\pm$ 0.13   & 5.0                    & 8.0                 & {[}-3, 3, 6, 12{]}                            & 17.5                         \\
       &                &                 & 0.170 $\pm$ 0.018 & 0.316 $\pm$ 0.030 & 1.2                    & 1.9                 & {[}-3, 3, 6{]}                                & 15.3                         \\
J1236  & 5.7            & 2.0             & 1.67 $\pm$ 0.10   & 1.75 $\pm$ 0.12   & 3.1                    & 5.3                 & {[}-3, 3, 6, 12, 24{]}                        & 58.3                         \\
       &                &                 & 0.552 $\pm$ 0.034 & 0.866 $\pm$ 0.056 & 1.5                    & 2.6                 & {[}-3, 3, 6, 12, 24{]}                        & 19.2                         \\
       &                &                 & 0.265 $\pm$ 0.019 & 0.564 $\pm$ 0.040 & 1.1                    & 1.9                 & {[}-3, 3, 6, 12{]}                            & 13.8                         \\
J1351  & 8.6            & 8.1             & 0.545 $\pm$ 0.063 & 3.48 $\pm$ 0.25   & 11.1                   & 19.0                & {[}-3, 3, 6{]}                                & 57.2                         \\
       &                &                 & -                 & -                 & -                      & -                   & -                                             & 13.6                         \\
       &                &                 & -                 & -                 & -                      & -                   & -                                             & 12.9                         \\
J1609  & 37.6           & 35.5            & 17.31 $\pm$ 0.87  & 30.6 $\pm$ 1.5    & 6.1                    & 4.2                 & {[}-3, 3, 6, 12, 24, 48, 96, 192{]}           & 84.9                         \\
       &                &                 & 5.31 $\pm$ 0.27   & 11.69 $\pm$ 0.59  & 2.4                    & 1.6                 & {[}-3, 3, 6, 12, 24, 48, 96, 192{]}           & 25.6                         \\
       &                &                 & 2.70 $\pm$ 0.14   & 7.27 $\pm$ 0.37   & 2.0                    & 1.4                 & {[}-3, 3, 6, 12, 24, 48, 96, 192{]}           & 13.8                    \\ \hline     
\end{tabular}
\end{table*}

\begin{figure}
    \centering
    \includegraphics[width=0.9\columnwidth]{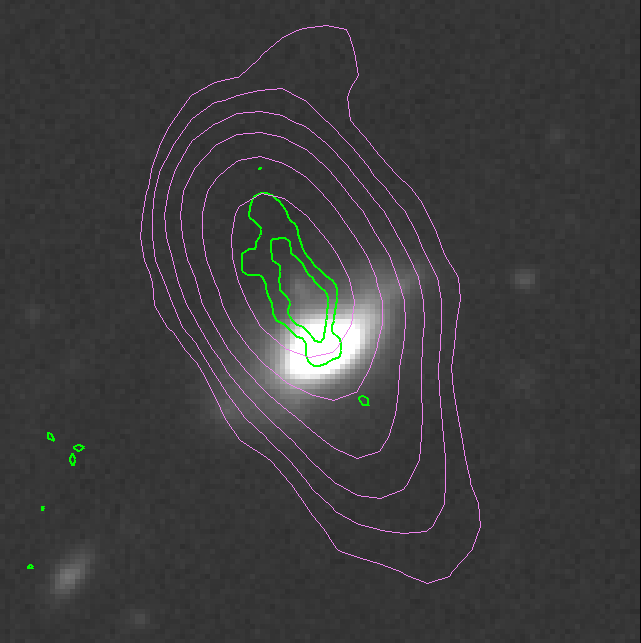}
    \caption{1.5 GHz contours from the VLA observations (green) and 144 MHz contours from LoTSS DR1 \citep[purple;][]{lotss_dr1} for the target J1351, overlaid on its optical (\textit{r}-band) image from \citet[][]{pierce19}. LoTSS contour levels begin at 5$\sigma_{\rm rms}$ and increase in factors of two up to 160$\sigma_{\rm rms}$, for $\sigma_{\rm rms}=54.5$ $\mu$Jy\,beam$^{-1}$. This target shows a diffuse, roughly linear radio emission structure at 1.5 GHz that propagates outwards from the optical centre on one side of the host galaxy. This structure appears to share a common centre with the LoTSS contours (limiting resolution $\sim$6"), which also provide evidence for additional emission on the opposite side of the galaxy. In combination, these features could be suggestive of remnant emission from a past phase of radio AGN activity.}
    \label{fig:J1351_opt}
\end{figure}

\textbf{J1108}. The extended emission for this source is confined to the scales of the galaxy bulge (up to $\sim$6 kpc in diameter), and exhibits a two-component structure in the C-band images that appears to be double-lobed, with a projected diameter of around 3 kpc. Although we consider this source to be double-lobed in nature, its appearance in the image at 7.5 GHz could also support a core-jet origin. Again, we do not see evidence for alignment with any optical structures.

\textbf{J12064}. This target exhibits very diffuse emission at all three frequencies, which is detected on projected scales that propagate to the galaxy halo in the optical image (up to around 8 kpc). Although present on large scales, the emission has no clear overall structure, and lacks the strong central emission that is seen for the majority of the extended sources. The galaxy does not show signs of strong star formation \citep[see Appendix in][]{pierce19}, and so this could indicate that this is remnant emission from a past phase of radio jet activity; see later description for target J1351.

\textbf{J1236}. This source has only marginally extended emission at 1.5 GHz, the structure of which is only slightly resolved in the C-band images; this is perhaps in part due to its relatively high redshift within the sample. Although it shows signs of symmetric structure in the North-West to South-East direction at 4.5 and 7.5 GHz that could be lobed, it is not possible to determine the true morphology with certainty from the current observations. Again, the radio emission appears to be confined to the projected scale of the host galaxy bulge ($\sim$5 kpc); the slight tail to the North-West in the L-band image is aligned with the appearance of the beam structure and is likely an artefact.

\textbf{J1351}. This target exhibits very diffuse emission with structure that is only clearly detected in its L-band image, but is not detected at the the higher frequencies due to the increased resolution. The structure at 1.5 GHz is the largest seen for the extended sources in the sample (diameter $\sim$19 kpc), and it is the only target for which the emission is clearly resolved in its FIRST image. Interestingly, the emission structure is not centred on its host galaxy, and in fact appears to propagate from its optical centre out into the surrounding galaxy environment to the North-East; the Southern edge of the VLA structure is centred on the optical nucleus of the galaxy, but there is no evidence for a compact radio core. This can be seen in Figure~\ref{fig:J1351_opt}, in which the 1.5 GHz contours from the VLA observations are overlaid on the optical image of the host galaxy. This target is also strongly detected at 144 MHz, with integrated flux $F_{\rm 144 MHz}\sim41$ mJy\,beam$^{-1}$, as determined from the first data release of the LOFAR Two-metre Sky Survey \citep[LoTSS;][]{lotss_dr1}. LoTSS contours are also overlaid on the optical image in Figure~\ref{fig:J1351_opt} (limiting resolution $\sim$6 arcseconds), appearing to share a common centre with the VLA contours and also showing evidence for additional emission on the opposing side of the galaxy; the same morphology is also seen when the FIRST contours are overlaid. In combination, these results could suggest that the radio jets have temporarily switched off in this object, leaving only remnant radio emission from a previous phase of radio AGN activity -- see \citet{mor17} for a review of radio AGN life cycles. Although we favour this interpretation, we cannot rule out that the emission structure belongs to a larger, more distant radio source; indeed, both sets of radio contours appear to centre on a patch of optical emission to the North-East, which could in fact be a background galaxy. 

\textbf{J1609}. This source shows a clear double-lobed structure in its 4.5 and 7.5 GHz images, with projected physical diameter $\sim$1--2 kpc. The emission is only extended on the scale of the bulge of its host galaxy (maximum diameter $\sim$4 kpc), although the lobes are aligned with the plane of its large, edge-on disk. 

Looking at the extended sources as a whole, there is therefore no favoured morphological structure amongst the objects that exhibit extended radio emission. Two galaxies, J12064 and J1351, display notably diffuse emission structures at all three frequencies and appear to lack prominent emission on small scales (e.g. compact radio cores). Although this type of emission structure is more typical of star forming regions than AGN jets, all targets in the sample were found to lack the far-infrared emission that would be expected if star formation contributed significantly to the radio emission \citep[see Appendix in][]{pierce19}.
These structures, however, could also be indicative of remnant radio emission from a past phase of activity \citep{mor17}. As discussed above, J1351, in particular, shows strong evidence for this latter interpretation.

The remaining 5 extended sources have collimated morphologies that are more representative of those expected to be produced by AGN jets. Of these, three show evidence for double-lobed morphologies with projected diameters (5$\sigma$ boundaries; see below) of $\sim$1--3 kpc: J0836 (at 7.5GHz); J1108 (at 4.5 GHz and 7.5 GHz); and J1609 (at 4.5 GHz at 7.5 GHz) -- these sizes do not account for the effects of beam smearing, however, so we also note that the distances between the peak fluxes of the two lobes in these cases are of order 0.5, 0.8 and 1.3 kpc, respectively. J1236 also exhibits a roughly symmetric structure that shows hints of being lobed, although this is not resolved well enough to be characterised with certainty. The other galaxy in this group, J0902, shows a morphology that is more reminiscent of FRI radio sources, with a highly collimated, jetted structure close to the nucleus (central $\sim$2 kpc) and a more diffuse lobes on larger scales that contribute to an overall `S-shaped' structure ($\sim$9--16 kpc).

The maximum angular extent of the emission structures for the extended sources were measured using 5$\sigma$ contours as limiting boundaries for the emission, which were subsequently converted to estimates of their maximum projected physical diameters. The emission structures range from 1.4 to 19.0 kpc in their maximum projected extents, with median values of 6.1 kpc, 3.7 kpc and 2.2 kpc at 1.5 GHz, 4.5 GHz and 7.5 GHz, respectively. 
We note that caution should be taken when interpreting these measurements, however, since the diameter estimates have not been corrected for beam-smearing effects, and the median physical scales of the beams at the three frequencies are 2.4 kpc (1.5 GHz), 0.89 kpc (4.5 GHz) and 0.55 kpc (7.5 GHz; from averaging the two beam axes). Certainly, beam smearing could at least partially explain why the maximum diameters are larger at 1.5 GHz in all cases. The largest angular and projected physical scales of the emission structures for the extended targets are listed in Table~\ref{tab:ext_info}.

The 5$\sigma$ contours were also used to measure the total flux values for the extended sources, with the peak flux values representing the maximum flux per beam detected within these regions; these results are also presented in Table~\ref{tab:ext_info}. The errors on these measurements were calculated in the same way as described for the unresolved sources in \S\ref{sec:point}, although fitting errors were not relevant in these cases. The 1.5 GHz total fluxes are compared to the FIRST fluxes in Figure~\ref{fig:FIRSTvsVLA}. All values are seen to lie below the line of equality, and in this instance it is found that only 2 out of 7 of the sources lie within 3$\sigma$ of their FIRST values. Unsurprisingly, the discrepancy is notably larger for J12064 and J1351 (both $>$18$\sigma$), the two targets that show the most diffuse emission structures in the contour maps; only $\sim$40 per cent of the FIRST fluxes are recovered in each case. This suggests that a significant amount of flux has been resolved out due to the higher resolution of the VLA observations, and confirms that these sources are typically much less compact than the unresolved objects.

\section{Discussion}
\label{sec:disc}

Our high-resolution VLA observations of a sample of 16 radio-intermediate HERGs have revealed a variety of morphologies in their emission structures. For the majority of galaxies in the sample (9 out of 16; 56 per cent) the morphology of the emission is unresolved at a limiting angular resolution of $\sim$0.3 arcseconds, corresponding to maximum projected physical diameters of $\sim$280--540 pc for these sources. For the remaining galaxies (7 out of 16; 44 per cent), the emission is extended in the images at all three frequencies, with projected maximum physical extents ranging from 1.4--19.0 kpc. There is no favoured morphological type within this group, and only 19 per cent of the sample (3 out of 16) show clear evidence for double-lobed emission structures at the frequencies considered. In this section, these results are discussed and compared with those found for FR0s, radio-intermediate type 2 quasars, and radio-powerful HERGs and Seyfert galaxies.

\subsection{Relation to FR0s}
\label{sec:FR0_comp}

Several studies have now shown that the majority of local radio AGNs have unresolved radio structures at the resolution limits of typical radio surveys \citep[e.g.][]{baldi10a,sad14,whit16}. These sources are often collectively named `FR0s', to distinguish them from the two traditional \citet[][]{fr74} classes of extended radio structures: FRIs (edge-darkened) and FRIIs (edge-brightened). High-resolution VLA observations of these compact AGN radio sources at frequencies of 1.5 GHz, 4.5 GHz and 7.5 GHz have revealed that the majority (22 out of 29; 76 per cent) remain unresolved or marginally resolved at limiting angular resolutions of order 0.3 arcseconds, corresponding to projected physical scales of $\sim$100--400 pc \citep{baldi15,baldi19a}. Given that we find that 56 per cent of our sample have unresolved radio emission structures at the same limiting angular resolution and similar projected physical scales ($\sim$280--540 pc) when using an identical observing strategy, it is natural to compare the properties of these objects with those of FR0 sources. 

Firstly, we note that although FR0s are predominantly associated with low-excitation optical emission -- recent studies even restrict their definitions of FR0s to only those with LERG spectra \citep[e.g.][]{baldi15,baldi18b,baldi19a}, which we denote as LERG/FR0s from this point forward -- many HERG sources also have compact radio morphologies \citep[e.g.][]{baldi10a}. However, the only compact HERG source that has previously been observed at high-resolution at 1.5, 4.5 and 7.5 GHz exhibits extended radio emission in its C-band images \citep[4.5 and 7.5 GHz;][]{baldi15}. This source has intermediate radio power (log(L$_{\rm 1.4GHz}$) $= 23.35$ W\,Hz$^{-1}$), and shows a clear double-lobed/edge-brightened structure of $\sim$0.8 kpc in diameter.\footnote[4]{Another higher radio power (log(L$_{\rm 1.4GHz}$) $= 25.03$ W\,Hz$^{-1}$) HERG source was also observed by \citet[][]{baldi15}, but this is is clearly extended in its FIRST image and shows a complex, potentially multiple-lobed morphology on large scales (up to $\sim$40 kpc).}

The remaining sources in the \citet{baldi15,baldi19a} samples are compact LERGs, and only a minority show extended radio structures in the VLA images. Most of these (5 out of 6) have strong and clearly visible core components at high resolution and, as a result, these radio structures tend to be edge-darkened. We do not see clearly detected cores for any of our radio-intermediate HERGs with extended radio structures, which in fact also show a preference for edge-brightened or diffuse emission (see Figure~\ref{fig:cont_a}). The detection of strong core components for the LERG/FR0s also enabled their central radio spectra to be easily determined, regardless of whether or not the sources showed extended emission, and we now compare these spectra with those obtained for the unresolved sources in our sample. We focus on the \citet[][]{baldi19a} sample for this comparison, since these objects were selected from the same \citet{bh12} parent sample as our radio-intermediate HERGs.

\begin{figure}
    \centering
    \includegraphics[width=0.95\columnwidth]{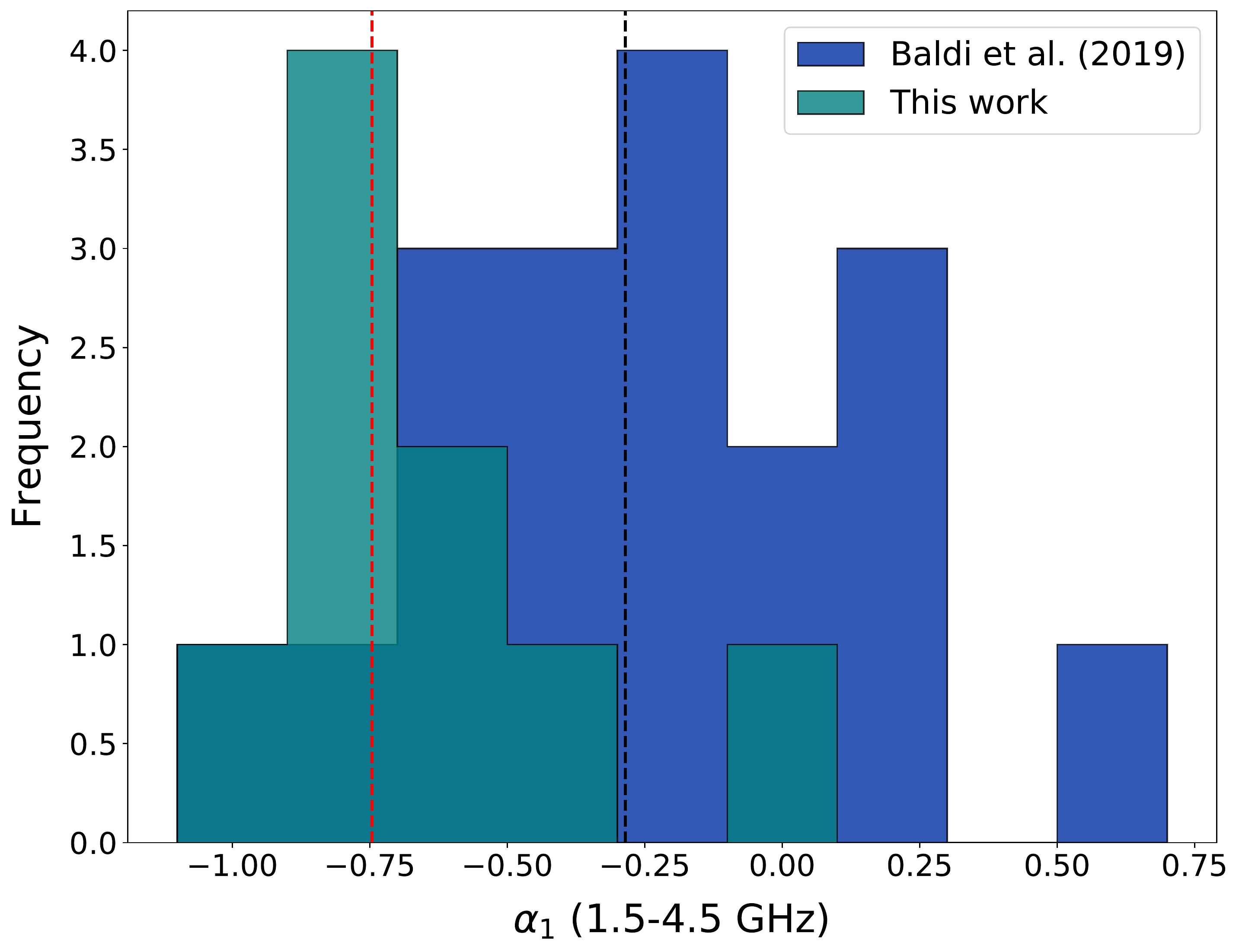}
    \caption{The distributions of the spectral indices between 1.5 and 4.5 GHz for the unresolved sources in our radio-intermediate HERG sample (teal) and all 18 LERG/FR0s studied in \citet[][; blue]{baldi19a}. The median values for the two samples are indicated by the red and black dashed lines, respectively.}
    \label{fig:spec_hist}
\end{figure}

Although the unresolved objects in our HERG sample (10$^{23}$ $<$ L$_{\rm 1.4GHz}$ $<$ 10$^{24}$ W\,Hz$^{-1}$) have similar radio powers to LERG/FR0s \citep[10$^{22}$ $\lesssim$ L$_{\rm 1.4GHz}$ $\lesssim$ 10$^{24}$ W\,Hz$^{-1}$;][]{baldi18a}, the typical radio spectra in the range 1.5 to 7.5 GHz appear to differ between the two populations. \citet[][]{baldi19a} found that 11 of the 18 LERG/FR0s they observed have flat spectra between 1.5 GHz and 4.5 GHz (61 per cent), defined as having spectral indices (for $F_{\nu}\propto\nu^{\alpha}$) in the range $-$0.4 $<$ $\alpha$ $<$ 0.2; the median value for the sample as a whole is $\alpha=-0.29$. Only one out of the 9 unresolved HERG objects studied here (11 per cent), the source J1358, has a spectral index that lies within this range when measured in the same way, with the other sources displaying steep spectra ($-$0.99 $<$ $\alpha_1$ $<$ $-$0.47); median $\alpha_1=-0.75$. A two-sample Kolmogorov-Smirnov test highlights the difference, indicating that the null hypothesis that the 9 unresolved radio-intermediate HERG sources and 18 \citet{baldi19a} LERG/FR0s are drawn from the same underlying distribution can be rejected at a confidence level of 98.3 per cent ($p=0.017$). The distributions of the spectral indices between 1.5 and 4.5 GHz of these two samples is presented in Figure~\ref{fig:spec_hist}, which illustrates this difference. 

Interestingly, \citet[][]{baldi19a} observed a spectral steepening for their objects between 4.5 GHz and 7.5 GHz, which is also observed for the unresolved sources here (see Figure~\ref{fig:spec_steepening}). We find that the steepening is in fact stronger, however, for the unresolved radio-intermediate HERGs; the median difference in spectral index is determined to be $\Delta\alpha = 0.33$, compared with the value of $\Delta\alpha = 0.16$ measured by \citet[][]{baldi19a}. One caveat that could affect this result, as well as those above, is that the LERG/FR0s in the \citet[][]{baldi19a} sample typically have lower redshifts (all $z<0.05$) than those of our radio-intermediate HERGs ($0.03<z<0.1$, median $z=0.082$), meaning that our measured fluxes probe the spectra at higher rest-frame frequencies in most cases. However, we note that this is likely to be a relatively minor effect.

Further differences between the two samples become apparent when considering their general host galaxy properties. LERG/FR0s are typically characterised by luminous, red early-type host galaxies, large black hole masses (typically $10^8$ $\lesssim$ $M_{\rm BH}$ $\lesssim$ $10^9$ $M_{\odot}$) and, of course, low-excitation optical emission \citep[][]{baldi18a}. They share most of these properties with the hosts of traditional FRI sources that generally exist at higher radio powers \citep[e.g. 10$^{23}$ $\lesssim$ L$_{\rm 1.4GHz}$ $\lesssim$ 10$^{25}$ W\,Hz$^{-1}$ in the FRI\textit{CAT} of][]{cap17}, with the only other clear difference lying in the compactness of the FR0 radio structures.

In contrast, the unresolved objects in our HERG sample have black hole masses that show a preference for lower values, lying in the range $10^{7.4}$ $\lesssim$ $M_{\rm BH}$ $\lesssim$ $10^{8.3} M_{\odot}$, with a median of $M_{\rm BH} = 10^{7.8} M_{\odot}$\footnote[5]{These were estimated from the velocity dispersion measurements ($\sigma_{*}$) provided by the MPA/JHU value-added catalogue for SDSS DR7 galaxies (available at: \url{https://wwwmpa.mpa-garching.mpg.de/SDSS/DR7/}), using the formulation of log$(M_{\rm BH}/M_{\odot}) = 8.13 + 4.02$ log$(\sigma_{*}/200$\,km\,s$^{-1}$) from \citet{trem02}.}. Furthermore, \citet{pierce19} showed that the host galaxies of radio-intermediate HERGs have a variety of optical morphologies, from late-types through to early-types, as identified using both visual inspection and light profile modelling. This is reflected by the unresolved HERG sources within the current sub-sample, but with a slight preference for early-type galaxies relative to late-types; a ratio of 2:4:3 is seen for late-type\,:\,early-type\,:\,mergers. We note, however, that the contrast between the host properties and black hole masses of our unresolved HERGs and the \citet[][]{baldi19a} LERG/FR0s is perhaps unsurprising, since these differences are apparent for the general populations of LERGs and HERGs from which these samples have both been drawn \citep[][]{bh12}.

In summary, the sources in our radio-intermediate HERG sample therefore have radio morphologies, radio spectra (for the unresolved objects) and host/nuclear properties that are significantly different from those of LERG/FR0 sources, and consequently also those of FRI sources at higher radio powers. Hence, it appears that they are representative of an important population of compact radio AGN sources that is distinct from those typically classified as FR0s and that has not yet been explored in detail.

\subsection{Comparison with radio-intermediate type 2 quasars}
\label{sec:T2_comp}

Recently, \citet{jar19} studied a sample of 10 local type 2 quasars with intermediate radio powers ($z<0.2$; 23.3 $<$ log(L$_{\rm 1.4GHz}$) $< 24.4$ W\,Hz$^{-1}$), which were also observed with the VLA in the L- and C-bands at a similar limiting angular resolution to our observations ($\sim$0.2--0.3 arcseconds). These objects are comparable in radio power to our radio-intermediate HERGs, but differ mainly in their much higher optical luminosities -- all of the \citet{pierce19} HERGs have [OIII]\,$\lambda$5007 luminosities below the 10$^{42}$ erg\,s$^{-1}$ lower limit for the \citet{jar19} sample. The type 2 quasars were also originally selected to have luminous, broad [OIII]\,$\lambda$5007 line components, i.e. which contribute $\geq$ 30 per cent of the total line flux and have a FWHM exceeding 700 km\,s$^{-1}$.

For the 9 \citet{jar19} objects for which radio AGN jet activity is found to dominate the total radio emission, it is seen that a much higher proportion (8 out of 9; 89 per cent) exhibit extended emission structures than for our radio-intermediate HERGs (44 per cent). Like the extended objects in our sample, the morphologies of the radio structures for the type 2 quasars are varied, although the latter are typically extended on larger scales. For instance, the median largest linear scale for the radio structures of the type 2 quasars is 8.3 kpc, whilst the median maximum projected extent (in the L-band images) for our extended radio-intermediate HERG sources is 6.1 kpc. One caveat with this comparison is that our maximum projected diameters were measured using 5$\sigma$ contours as the boundaries for the emission (see \S\ref{sec:ext}), whilst the largest linear scales for the \citet{jar19} objects were measured between the peaks of the two furthest separated radio structures; the radio structures seen for our extended HERG sources typically do not allow for this latter technique. However, since this means that our measurements are more susceptible to beam smearing, we stress that this only serves to increase the discrepancy in the extension of the emission between the two samples.

Although the \citet{jar19} sources display a variety of radio morphologies, two show clear edge-brightened/double-lobed structures in their high-resolution C-band images (median resolution $\sim$0.2 arcseconds), with largest linear scales of $\sim$0.9 and 1.4 kpc. This is comparable with the scale of the similar structures found for both the compact HERG source discussed in \S\ref{sec:FR0_comp} ($\sim$0.8 kpc) and for the examples in our current sample ($\sim$0.5--1.3 kpc in peak-to-peak distance). Interestingly, these sizes are intermediate between those of similar structures in Seyfert galaxies \citep[typically few 100 pc][]{uw84} and in radio-powerful HERG/FRII sources (few 10s kpc, up to Mpc scales). We compare the results for our extended sources with these populations in the following section.

\subsection{Comparison with radio-powerful HERGs and Seyferts}
\label{sec:HERG_comp}

As mentioned in \S\ref{sec:int}, many studies have suggested that high-excitation optical emission from AGNs is linked with a radiatively efficient black hole accretion mode, regardless of the level of associated radio emission \citep[e.g.][]{bh12,hb14,min14}. At high radio powers, it is seen that the vast majority of strong-line radio galaxies (SLRGs)\footnote[6]{This is a class that overlaps significantly with HERGs but is defined solely on the basis of [OIII]$\lambda$\,5007 equivalent width.} are associated with classical FRII  radio morphologies \citep{tad16}. At low radio powers, high-resolution radio observations of nearby Seyfert galaxies have revealed that many also display double-/triple-lobed or edge-brightened \citep{uw84,mor99,baldi18b} radio structures, although on much smaller scales (typically $\lesssim$400 pc). 

This could suggest that HERGs and Seyfert galaxies in fact compose a continuous AGN population with a common black hole accretion mode that leads to edge-brightened, double-lobed radio structures, with the jet-launching mechanism perhaps providing the link between the two. In this picture, radio-intermediate HERGs would represent the missing population between radio-powerful HERGs at high radio powers and Seyfert galaxies at low radio powers, and should therefore also preferentially display multiple-lobed or edge-brightened extended radio structures. 

We have found, however, that only 3 out of the 16 galaxies in the sample (19 per cent) appear to show evidence for possible edge-brightened, double-lobed structures in at least one of the images at the three observation frequencies. On the other hand, it is important to consider that 9 out of the 16 galaxies in the sample have unresolved radio structures at all three frequencies, for which the upper limits on the projected physical diameters ($d_{\rm max}$ in Table~\ref{tab:ps_fluxes}) range from $\sim$280--540 pc, with a median of 470 pc. Given that the edge-brightened structures seen for some Seyfert galaxies typically manifest on scales below $\sim$400 pc, and can even appear on sub-100 pc scales \citep[e.g. as seen for Mrk 348;][]{nd83}, we cannot definitively rule out that such radio structures still exist within these galaxies. 

Looking at the extended sources alone, we find that a slight minority (3 out of 7; 43 per cent) show some suggestion of double-lobed structures. Two of the remaining galaxies (J12064 and J1351), however, display diffuse emission structures in the images and lack strong emission on small scales. Although this could suggest that star formation provides the main contribution to their radio emission, this was deemed to be unlikely for both galaxies based on the properties of their optical and far-infrared emission \citep[see Appendix in][]{pierce19}. As discussed in \S\ref{sec:ext}, we therefore favour the explanation that the radio jets have temporarily switched off in these objects to leave only remnant radio emission from a previous phase of radio AGN activity \citep[e.g. see][]{mor17}; the case is particularly strong for J1351, which also has a strong detection at 144 MHz in LoTSS DR1 \citep{lotss_dr1}. 

Of the remainder, we find that J1236 shows a roughly symmetric emission structure on scales of up to $\sim$5 kpc that could also be lobed, although it is not resolved well enough to determine this with certainty. J0902 also shows clear jet-related structure on small scales, but which appears in an `S-shape', with `tails' of emission that diverge from the main axis of the jet around 2--3 kpc from the centre. Therefore, considering the sources that show strong evidence of currently being in a phase of activity in which the radio jets are switched on, we find that the majority (3 out of 5; 60 per cent) exhibit double-lobed structures. 

Given that the unresolved sources in the sample show differences from the properties of typical LERG/FR0 sources and could still show double-lobed/edge-brightened structures on smaller physical scales than those probed here, the results could therefore still be consistent with the picture of a continuous connection between radiatively-efficient accretion and edge-brightened radio emission structures along the HERG sequence, for currently active jets. Clearly, high-resolution radio observations of greater numbers of radio-intermediate HERGs are required to test this idea further.
Ideally, such observations would be done at higher resolution than those used here, in order to ensure that the structures are investigated in appropriate detail, including the unresolved HERG sources in our current sample.

\section{Summary and Conclusions}
\label{sec:sum}

In the past decade, high sensitivity, large area radio surveys have revealed that compact sources with low-to-moderate radio powers dominate the local radio AGN population. Local HERGs with intermediate radio powers (22.5 $<$ log(L$_{\rm 1.4GHz}$) $< 25.0$ W\,Hz$^{-1}$) form an important sub-population that also shows strong evidence for driving outflows on the scale of galaxy bulges. Here, we have used high-resolution radio observations of a sample of 16 radio-intermediate HERGs at 1.5 GHz, 4.5 GHz and 7.5 GHz to characterise the morphologies and extents of their radio structures down to sub-kpc scales. Our main results are as follows.

\begin{itemize}
\item The majority of targets in the sample (9 out of 16; 56 per cent) are unresolved at an angular resolution of $\sim$0.3 arcseconds, placing an upper limit on their projected physical diameters of $\sim$280--540 pc. All of these sources show steep radio spectra between 1.5 GHz and 4.5 GHz ($-$0.99 $<$ $\alpha$ $<$ $-$0.47, for $F_{\nu}\propto\nu^{\alpha}$), except one (J1358), which has a flat spectrum ($\alpha_1=-0.05 \pm 0.01$). The latter also has a 1.5 GHz flux that is greatly discrepant from its FIRST value, which could suggest it is a beamed, variable source. All of this group, including J1358, show a systematic spectral steepening between 4.5 GHz and 7.5 GHz (median $\Delta\alpha=0.33$). 

\item 7 sources in the sample display extended emission structures with projected diameters of 1.4--19.0 kpc. Of these, it is found that three (43 per cent) show suggestions of double-lobed emission structures in at least one of the images at the three frequencies used. One other extended source shows signs of a symmetric emission structure that could be lobed, but it is not resolved well enough to classify with certainty. Of the remainder, two sources (J12064 and J1351) show diffuse emission structures in the images that could be suggestive of remnant emission from past phases of jet activity. The remaining source (J0902) shows an `S-shaped' morphology that could be suggestive of either an interaction between an inclined jet and a disk of dense gas, or jet precession. The majority of the sources with clear evidence for currently active radio jets therefore exhibit double-lobed structures (3 out of 5; 60 per cent).

\item Although compact on similar scales, the radio-intermediate HERG sources are found to differ from LERG/FR0s in their radio morphologies, radio spectra (for the unresolved objects) and, as found in previous studies of larger HERG and LERG samples, their host properties and black hole masses.
\end{itemize}

These results provide some evidence to suggest that there is a link between double-lobed/edge-brightened radio structures and the radiatively-efficient accretion mode often associated with HERGs, although the strength of this connection remains uncertain. However, they also reveal that a significant population of compact AGN radio sources with high-excitation optical emission exists, which up until now has not been explored in detail. Additional high-resolution radio observations of HERGs with low-to-intermediate radio powers are therefore required to improve our understanding of this population, and further investigate the link between AGN accretion mode and radio morphology. This includes follow-up radio observations of the unresolved sources in the current sample at higher angular resolutions, to rule out the possibility that double-lobed structures exist on the scales at which they can manifest in Seyfert galaxies.

\section*{Acknowledgements}

The authors wish to thank the referee for their constructive comments and suggestions, which helped to improve the overall quality of the paper. JP thanks R. Baldi for helpful discussions. Both JP and CT acknowledge support from STFC funding. The National Radio Astronomy Observatory (NRAO) is a facility of the National Science Foundation operated under cooperative agreement by Associated Universities, Inc.




\bibliographystyle{mnras}
\bibliography{master} 



\appendix

\section{Observations of HERGs with lower radio powers}
\label{sec:app}

\begin{table*}
\caption{Basic information for the two HERG sources with lower radio powers, which were observed in addition to those in the main sample. Columns are mostly as in Table~\ref{tab:target_info}, although in this instance the dates of the target observations are also listed in the final column.}
\label{tab:app_target_info}
\begin{tabular}{cccccccc}
\hline
SDSS ID                       & z     & \makecell{log(L$_{\rm 1.4GHz}$)\\{[}W\,Hz$^{-1}${]}} & \makecell{log(L$_{\rm O[III]}$)\\{[}W{]}} & log(M$_{*}$/M$_{\odot}$) & log(M$_{\rm BH}$/M$_{\odot}$) & Optical morph. & Obs. date  \\ \hline
J091107.04$+$454322.6 (J0911) & 0.098 & 22.62                                                   & 33.26                         & 10.7                     & 7.5                           & Spiral (L)     & 2018-03-28 \\
J101207.79$+$084235.1 (J1012) & 0.095 & 22.39                                                   & 33.81                         & 10.5                     & 7.7                           & Spiral (L)     & 2018-03-28
\\ \hline
\end{tabular}
\end{table*}

\begin{figure*}
    \centering
    \includegraphics[width=\textwidth]{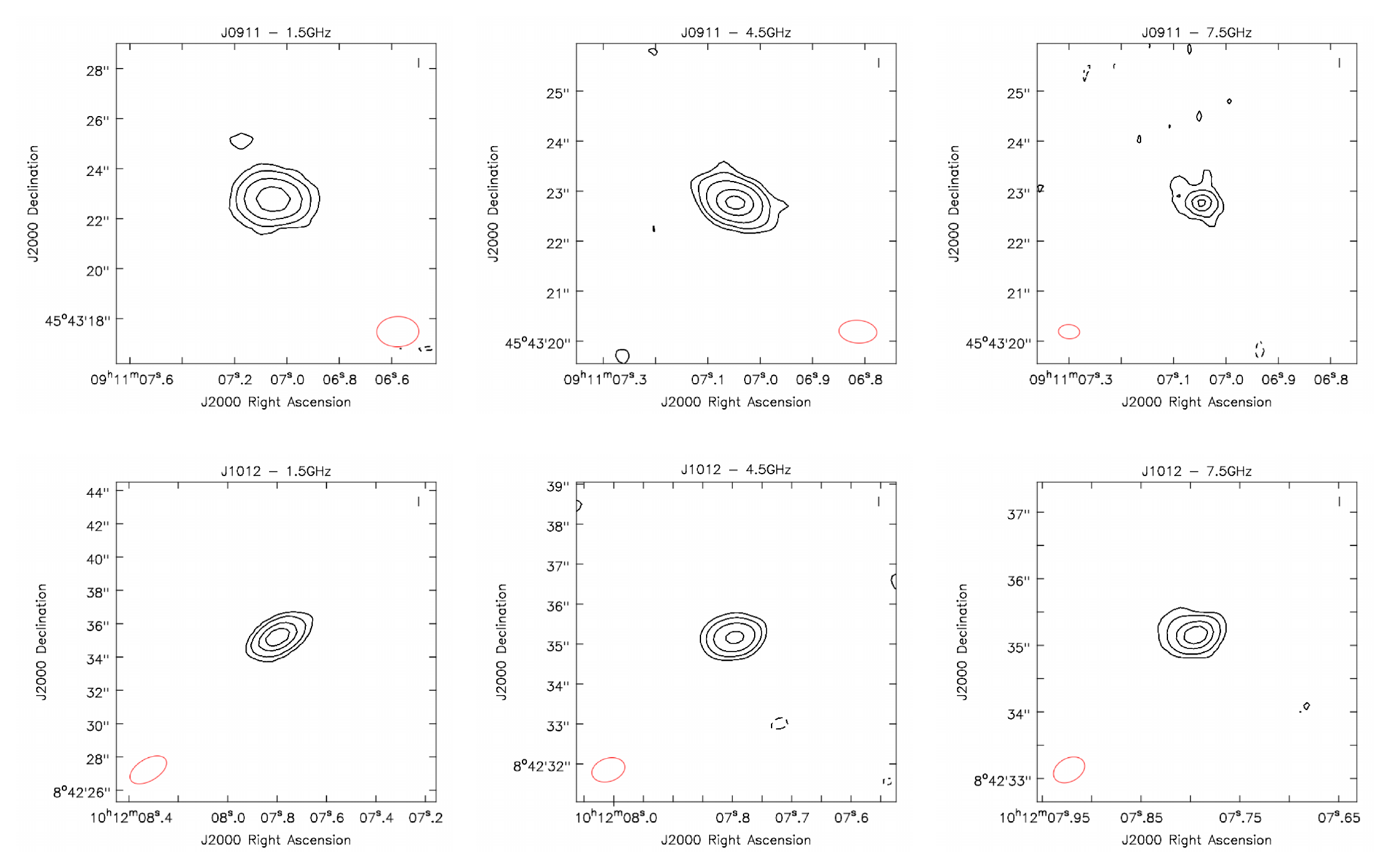}
    \caption{Contour plots for the two HERGs with lower radio powers at 1.5 GHz (left column), 4.5 GHz (middle column) and 7.5 GHz (right column). Positive contours (solid lines) begin at $3\sigma$ and typically increase by subsequent factors of two; exact contour levels are listed in Table~\ref{tab:cont_app}. Negative contours at the $-3\sigma$ level are indicated with dashed lines. The red ellipses indicate the approximate beam shape for each image.}
    \label{fig:cont_app}
\end{figure*}

In addition to the sample of 16 radio-intermediate HERGs presented in the main body of the paper, two other HERG sources with lower radio powers [log(L$_{\rm 1.4GHz}$) $=$ 22.39 and 22.62] were also observed with the VLA in the same period using the same observing strategy. The data reduction and image production were performed in an identical manner to that carried out for the targets in the main sample, as outlined in \S\ref{sec:obs}. Basic information on these targets is presented in Table~\ref{tab:app_target_info}. The contour maps for these sources are presented in Figure~\ref{fig:cont_app}, and the contour levels are presented in Table~\ref{tab:cont_app}. The nature of the emission for these targets and their relation to the optical appearance of the host galaxies is briefly described below.

\textbf{J0911.} This target has slightly extended structure in the images at 4.5 GHz and 7.5 GHz, although this is only just resolved at 1.5 GHz; there is a patch of emission above the 3$\sigma$ threshold in the 1.5 GHz image, but there is little difference between the peak and total flux values for this source (see Table~\ref{tab:cont_app}). The emission is confined to the scales of the bulge of its host galaxy and does not align with any optical features. Given that the emission structure is only just resolved, it is not possible to characterise its appearance further.

\textbf{J1012.} This source only has clearly resolved emission in its image at 7.5 GHz, although it appears to be marginally extended along the same position angle in its 4.5 GHz image. As with J0911, the emission is confined to the host galaxy bulge and shows no alignment with optical features. 

Flux measurements and error calculations for the sources were conducted using the same methods as outlined in \S\ref{sec:analysis}, and are listed in Table~\ref{tab:cont_app}. Both sources are found to have fluxes at 1.5 GHz that exceed their FIRST values, although for J0911 the fluxes are consistent when considering the errors on the VLA measurements. For J1012, however, the difference appears to be more significant (at the 4.9$\sigma$ level). The fact that the VLA flux is higher than the FIRST value means that this cannot be caused by flux losses introduced by the improved resolution of the VLA observations, and the $\sim$30 per cent discrepancy appears to be too large to be explained by variability. The cause of this difference is therefore uncertain. 

Given their relatively high redshifts within the sample, higher-resolution observations are important for determining the radio structures of these sources in more detail, particularly because both galaxies show signs of extended emission, at some level.

\begin{table*}
\centering
\caption{Emission properties and contour levels for the continuum maps displayed in Figure~\ref{fig:cont_app}. The three rows for each target indicate the values for the 1.5 GHz, 4.5 GHz and 7.5 GHz maps, respectively. The beam parameters are presented as in Table~\ref{tab:image_prop}, and the peak fluxes ($f_{\nu}^{P}$) and total fluxes ($f_{\nu}^{T}$) as in Table~\ref{tab:ext_info}. FIRST and NVSS fluxes for the targets are also listed for comparison. The maximum angular ($\theta_{\nu}^{\rm max}$) and projected physical extents ($d_{\nu}^{\rm max}$) are also as presented in Table~\ref{tab:ext_info}, although the bracketed values for J1012 at 1.5 GHz are calculated from the average of the two beam axis lengths, as in Table~\ref{tab:ps_fluxes}. Contour levels are presented as multiples of the background rms measurements in the images ($\rm \sigma_{rms}$), the values of which are also listed in the final column. The units are presented in brackets beneath their associated measurements, where appropriate.}
\label{tab:cont_app}
\begin{tabular}{ccccccccccc}
\hline
Name  & \makecell{F$_{\rm NVSS}$\\(mJy)} & \makecell{F$_{\rm FIRST}$\\(mJy)} & \makecell{$f_{\nu}^{P}$\\(mJy)}     & \makecell{$f_{\nu}^{T}$\\(mJy)}     & \makecell{$\phi_{\nu}^{\rm max}$\\(arcsec)} & \makecell{$d_{\nu}^{\rm max}$\\(kpc)} & \makecell{$\rm \theta_{1.5}$\\(arcsec)} & \makecell{$\rm \psi_{1.5}$\\(deg)} & \makecell{Contour levels\\($n\sigma_{\rm rms}$)} & \makecell{$\sigma_{\rm rms}$\\($\mu$Jy)} \\ \hline
J0911 & 8.6            & 1.9             & 1.92 $\pm$ 0.11   & 1.98 $\pm$ 0.13   & 3.2                    & 5.5                 & 1.68 $\times$ 1.21                                                                     & -88.8                                                                             & {[}-3,3,6,12,24{]}                                                                       & 54.5                                                                                    \\
      &                &                 & 0.670 $\pm$ 0.037 & 0.882 $\pm$ 0.053 & 1.5                    & 2.6                 & 0.76 $\times$ 0.45                                                                     & 86.2                                                                              & {[}-3,3,6,12,24,36{]}                                                                       & 15.7                                                                                    \\
      &                &                 & 0.346 $\pm$ 0.025 & 0.448 $\pm$ 0.036 & 0.7                    & 1.3                 & 0.42 $\times$ 0.28                                                                     & 86.5                                                                              & {[}-3,3,6,12,18{]}                                                                           & 18.0                                                                                    \\
J1012 & 11.1           & 1.2             & 1.83 $\pm$ 0.12   & -                 & (1.9)                  & (3.2)               & 2.43 $\times$ 1.31                                                                     & -59.2                                                                             & {[}-3,3,6,12,18{]}                                                                           & 74.8                                                                                    \\
      &                &                 & 0.761 $\pm$ 0.046 & 0.779 $\pm$ 0.055 & 1.5                    & 2.5                 & 0.84 $\times$ 0.58                                                                     & -71.1                                                                             & {[}-3,3,6,12,24{]}                                                                       & 26.5                                                                                    \\
      &                &                 & 0.375 $\pm$ 0.025 & 0.428 $\pm$ 0.032 & 0.9                    & 1.5                 & 0.50 $\times$ 0.35                                                                     & -60.9                                                                             & {[}-3,3,6,12,18{]}                                                                           & 15.8                     \\                  
       \hline
\end{tabular}
\end{table*}


\bsp	
\label{lastpage}
\end{document}